\documentclass{aa}
\usepackage[varg]{txfonts}

\begin{document}

   \title{Pebble drift and planetesimal formation in protoplanetary discs with embedded planets}

   \author{Linn E.J. Eriksson
          \inst{1}
          ,
          Anders Johansen\inst{1}
          , 
          Beibei Liu\inst{1}
          }

   \authorrunning{L.E.J. Eriksson}
   \institute{Lund Observatory, Department of Astronomy and Theoretical Physics, Lund University, Box 43, 221 00 Lund, Sweden\\
              \email{linn@astro.lu.se}}

   \date{Received X; accepted X}

\abstract{Nearly-axisymmetric gaps and rings are commonly observed in protoplanetary discs. The leading theory regarding the origin of these patterns is that they are due to dust trapping at the edges of gas gaps induced by the gravitational torques from embedded planets. If the concentration of solids at the gap edges becomes high enough, it could potentially result in planetesimal formation by the streaming instability. We test this hypothesis by performing global 1-D simulations of dust evolution and planetesimal formation in a protoplanetary disc that is perturbed by multiple planets. We explore different combinations of particle sizes, disc parameters, and planetary masses, and find that planetesimals form in all these cases. We also compare the spatial distribution of pebbles from our simulations with protoplanetary disc observations. Planets larger than one pebble isolation mass catch drifting pebbles efficiently at the edge of their gas gaps, and depending on the efficiency of planetesimal formation at the gap edges, the protoplanetary disc transforms within a few 100,000 years to either a transition disc with a large inner hole devoid of dust or to a disc with narrow bright rings. For simulations with planetary masses lower than the pebble isolation mass, the outcome is a disc with a series of weak ring patterns but no strong depletion between the rings. Lowering the pebble size artificially to 100 micrometer-sized ``silt'', we find that regions between planets get depleted of their pebble mass on a longer time-scale of up to $0.5$ million years. These simulations also produce fewer planetesimals than in the nominal model with millimeter-sized particles and always have at least two rings of pebbles still visible after $1\, \textrm{Myr}$.} 

\maketitle 

\section{Introduction}
High spatial resolution dust continuum observations with the Atacama Large Millimeter Array (ALMA) have shown that concentric rings and gaps are common features in protoplanetary discs (see e.g. \citealt{ALMA2015,Andrews2016,Isella2016,Pinte2016}). Detections of rings and gaps are most common in millimeter continuum emission, which traces the population of millimeter-sized pebbles (e.g. \citealt{Clarke2018,Dipierro2018,Fedele2018,Long2018,Long2019}); however, the same patterns have also been observed in the distribution of micron-sized dust grains via scattered light observations (e.g. \citealt{Ginski2016,vanBoekel2017,Avenhaus2018}), and in the gas surface density via observations of molecular emission (e.g. \citealt{Isella2016,Fedele2017,Teague2017,Favre2018}). Since these features are seen in both the solid and the gas component of the disc, they have been interpreted as a signature of planet-disc interactions (e.g. \citealt{Pinilla2012,Dipierro2015,Favre2018,Fedele2018}). Numerical simulations have long predicted that massive planets will open gap(s) in the disc, creating local pressure maxima at the gap-edges where particles can become trapped (e.g. \citealt{PapaloizouLin1984,PaardekooperMellema2004,Dong2015}). This idea has gained relevance, as \citet{Dullemond2018} recently found evidence that at least some of the rings seen in observations are due to dust trapping in radial pressure bumps. Since the dust-to-gas ratios in these pressure bumps can become significantly higher than the global value, they are a likely site for planetesimal formation via the streaming instability (\citealt{YoudinGoodman2005,JohansenYoudin2007,BaiStone2010,YangJohansen2014}). This is the key process that we investigate in this work.

In this paper we explore observational consequences of the formation of gaps by embedded planets. Particularly, we investigate whether particle trapping at the edges of planetary gaps is efficient enough to trigger the streaming instability and result in the formation of planetesimals. To this end we use a dust evolution model including both radial drift, stirring and coagulation, and perform first principle calculations in one dimension over large spatial scales ($1-500\, \textrm{au}$) and long disc evolution lifetimes ($1\, \textrm{Myr}$). The main questions that we want to answer can be summarized as follows: 1) Do planetesimals form at the edges of planetary gaps; 2) If so, how efficient is this process and how does the efficiency vary with different disc and planet parameters; 3) What does the distribution of dust and pebbles look like for the different simulations; 4) How do these distributions compare with observations of protoplanetary discs?

We find that planetesimals indeed do form at the edges of planetary gaps. For millimeter-sized pebbles and planetary masses larger than the pebble isolation mass, essentially all pebbles trapped at the pressure bump are turned into planetesimals. In combination with fast radial drift, this results in that the region interior to the outermost planet gets depleted of pebbles, leaving us with something that resembles a transition disc \citep{Andrews2011}. When the particle size is lowered to 100 $\mu$m, a larger dust-to-gas ratio is required to trigger the streaming instability. Because of this the planetesimal formation efficiency drops, and at least one ring of pebbles remain visible in the otherwise empty inner disc. When the planetary mass is lowered to less than the pebble isolation mass, trapping at the gap-edges becomes less efficient. Pebbles are also able to partially drift through the planetary gaps, resulting in a continuous replenishing of pebbles to the inner disc, and the pebble distribution appears as a series of weak gaps and rings at the locations of the planets.

In the next section (section \ref{sec:theory}) we present our models for disc evolution, dust growth and planetesimal formation. The numerical set-up of the simulations is described in section \ref{sect: numerical methods}. In section \ref{sect: results} we present results for the nominal model and the parameter study. In the nominal simulation the maximum pebble size reached by coagulation is around one millimeter; results from simulations where the maximum grain size was artificially decreased to 100 $\mu$m are presented in section \ref{sect:100micron}. In section \ref{sec: observations} we compare our results to observations, and in section \ref{setc: shortcomings of the model} we discuss the shortcomings of the model. The most important results and findings are summarized in section \ref{setc: conclusion}. In Appendix \ref{Appendix: collision algorithm} we describe the method used to model particle collisions, and in Appendix \ref{Appendix: sec size dist} we show particle size distributions for some interesting simulations in the parameter study.

\section{Theory}\label{sec:theory}
In our model we use a one-dimensional gas disc that is evolving viscously. We apply planetary torques to the disc in order to simulate gap-opening by planets. For the evolution of dust particles we use a model containing both particle growth, stirring and radial drift. We further include a model for planetesimal formation where the conditions for forming planetesimals are derived from streaming instability simulations.

\subsection{Disc model}
The initial surface density profile of the disc is chosen to be that of a viscous accretion disc,
\begin{equation}
\Sigma=\frac{\dot{M}_0}{3\pi \nu} \exp \left[-\frac{R}{R_\textrm{out}}\right],
\end{equation}
where $\Sigma$ is the surface density of the gas, $\dot{M}_0$ is the initial disc accretion rate, $\nu$ is the kinematic viscosity of the disc, $R$ is the semimajor axis, and $R_{\rm out}$ is the location of the outer disc edge (e.g. \citealt{Pringle1981}).
The evolution of the surface density is solved using the standard one dimensional viscous evolution equation from \citet{LinPapaloizou1986} for a disc that is being perturbed by a planet,
\begin{equation}\label{eq:surface density evolution}
\frac{\partial \Sigma}{\partial t}=\frac{1}{R}\frac{\partial }{\partial R} \left[3R^{1/2}\frac{\partial }{\partial R}(\nu \Sigma R^{1/2}) - \frac{2 \Lambda \Sigma R^{3/2}}{(GM_{*})^{1/2}} \right].
\end{equation}
In the above equation $t$ is the time, $\Lambda$ is the torque density distribution, $G$ is the gravitational constant, and $M_*$ is the stellar mass. Equation \ref{eq:surface density evolution} is essentially the continuity equation in cylindrical coordinates,
\begin{equation}\label{eq:continuity equation}
\frac{\partial \Sigma}{\partial t}=\frac{1}{R}\frac{\partial }{\partial R} \left(\Sigma v_{\textrm{R}} R \right),
\end{equation}
where the radial velocity $v_{\textrm{R}}$ has two components which can be obtained from comparison of the two equations.
The kinematic viscosity is approximated using the alpha approach from \citet{ShakuraSunyaev1973}, 
\begin{equation}
\nu = \alpha_{\textrm{visc}} \Omega H^2,
\end{equation}
where $\alpha_{\textrm{visc}}$ is a parameter related to the efficiency of viscous transport, $\Omega  = (GM_*/R^3)^{1/2}$ is the Keplerian angular velocity, and $H$ is the scale height of the disc. The scale height is calculated as
\begin{equation}
H = \frac{c_{\textrm{s}}}{\Omega},
\end{equation}
where $c_{\textrm{s}}$ is the sound-speed,
\begin{equation}
c_{\textrm{s}} = \left(\frac{k_{\textrm{B}} T}{\mu m_{\textrm{H}}}\right)^{1/2}.
\end{equation}
In the above equation $k_{\textrm{B}}$ is the Bolzmann constant, $T$ is the temperature, $m_{\textrm{H}}$ is the mass of the hydrogen atom, and $\mu$ is the mean molecular weight, set to be 2.34 for a solar-composition mixture of hydrogen and helium \citep{Hayashi1981}. We use a fixed powerlaw structure for the temperature, 
\begin{equation}
T=T_{\textrm{const}} \times (R/{\rm AU})^{-\zeta},
\end{equation}
with a radial temperature gradient of $\zeta=3/7$ and a mid-plane temperature of $T_{\textrm{const}}=150\, \textrm{K}$ at $1\, \textrm{au}$ \citep{ChiangGoldreich1997}.

\subsection{Planetary torque}
The effect on the disc due to the planet is governed by the torque density distribution, $\Lambda$, here defined as the rate of angular momentum transfer from the planet to the disc per unit mass. For modelling of the torque density distribution we follow \citet{D'AngeloLubow2010},
\begin{equation}\label{eq:torqueDensity}
\Lambda = - F(x,\beta,\zeta) \Omega_a^2 a^2 q^2 \left(\frac{a}{H_a}\right)^4.
\end{equation}
In the above equation $F$ is a dimensionless function, $x=(R-a)/H_a$, $\beta$ and $\zeta$ are the negative radial gradients of surface density respectively temperature, and $q$ is the planet-to-star mass ratio. The subscript $a$ denotes the location of the planet. The analytic expression used for function $F$ is 
\begin{equation}\label{eq:function F}
\begin{split}
F(x,\beta,\zeta) = \left\{p_1 \exp \left[-\frac{(x+p_2)^2}{p_3^2}\right] + p_4\exp \left[-\frac{(x+p_5)^2}{p_6^2}\right]\right\} \\
\times \tanh(p_7 - p_8x),
\end{split}
\end{equation}
where the parameters $(p_1,\ldots,p_8)$ are provided as a fit to actual simuations. Table 1 of \citet{D'AngeloLubow2010} gives values for these parameters for a set of discrete values of $\beta$ and $\zeta$. As mentioned in the previous subsection we use a fixed radial temperature gradient of $\zeta=3/7$, and we choose to simplify the problem even further by also using a constant surface density gradient of $\beta=15/14$. For these values of $\beta$ and $\zeta$ the parameters $(p_1,\ldots,p_8)$ take on the values listed in table \ref{table: parameter p}.

\begin{table}
\caption{Values of the parameters $p_{\textrm{n}}$ in equation \ref{eq:function F} for $\beta=15/14$ and $\zeta=3/7$. The values are obtained from linear interpolation using Table 1 in \citet{D'AngeloLubow2010}.}
\label{table: parameter p}
\centering
\begin{tabular}{c c}          
\hline\hline
$p_{\textrm{n}}$ & Value \\
\hline
    $p_1$ & 0.029355 \\
    $p_2$ & 1.143998\\
    $p_3$ & 0.918121 \\
    $p_4$ & 0.042707 \\
    $p_5$ & 0.859193 \\
    $p_6$ & 1.110171 \\
    $p_7$ & -0.152072 \\
    $p_8$ & 3.632843 \\
\hline\hline
\end{tabular}
\end{table}

\subsection{Pebble isolation mass}
The pebble isolation mass ($M_{\textrm{iso}}$) is defined as the mass when the planet can perturb the local pressure gradient in the mid-plane enough to make it zero outside the gap, thus creating a pressure bump. Pebbles drifting inwards will be trapped at the pressure bump, resulting in a locally enhanced solid density just outside the orbit of the planet (e.g. \citealt{Lambrechts2014,Pinilla2016,Weber2018}). The planetary masses used in our simulations will always be in units of pebble isolation masses.

We use an analytical fitting formula for the pebble isolation mass which was derived by \citet{Bitsch2018} using 3-D hydrodynamical simulations of planet-disc interactions,
\begin{equation}\label{eq: mIso B18}
M_{\textrm{iso}} = 25\, \textrm{M}_{\oplus} \times f_{\textrm{fit}},
\end{equation}
and
\begin{equation}
f_{\textrm{fit}} = \left[\frac{H/R}{0.05}\right]^3 \left[0.34 \left(\frac{\log (\alpha_3)}{\log (\alpha_{\textrm{visc}})}\right)^4 + 0.66\right]\left[1-\frac{\frac{\partial \ln P}{\partial \ln R} + 2.5}{6}\right].
\end{equation}
In the above equation $\alpha_3=0.001$.

The pebble isolation mass is extremely dependent on the gap depth, which is known to vary between one dimensional and multidimensional simulations (e.g. \citealt{LinPapaloizou1986,Kanagawa2015,HallamPaardekooper2017}). In general, 1-D simulations produce narrower and deeper gaps than their higher dimensional analogues. This means that the mass required to reach a radial pressure gradient of zero is smaller in 1-D simulations than in 3-D simulations. In other words the pebble isolation masses derived by \citet{Bitsch2018} is significantly higher than the pebble isolation mass obtained in our simulations using the tabulated torques of \citet{D'AngeloLubow2010}. 

We performed our own 1-D simulations to calculate the pebble isolation mass, and found during comparison that the pebble isolation masses obtained in 1-D and 3-D simulations are related by a scalar factor which only seem to depend on $\alpha_{\textrm{visc}}$. Approximate values of this scalar, which we denote $k_{3\textrm{D}/1\textrm{D}}$, can be found in Table \ref{table: bert factor} for a range of values of $\alpha_{\textrm{visc}}$. As an example: if $\alpha_{\textrm{visc}}=0.01$ then the pebble isolation mass at $11.8\, \textrm{au}$ is $59.6\, \textrm{M}_{\oplus}$ according to equation \ref{eq: mIso B18} and using the values quoted for $T_{\textrm{const}}, \beta$ and $\zeta$. The planetary mass required to obtain a zero pressure gradient outside a planet orbiting at the same semimajor axis in our simulations is $59.6/1.5=39.7\, \textrm{M}_{\oplus}$. Figure 3 of \citet{Johansen2019} shows a similar systematic difference between the 1-D and the 3-D gap depth.

In our simulations, to avoid working with an artificially low pebble isolation mass due to the 1-D approach, we modify the magnitude of the torque density distribution in the following way,
\begin{equation}\label{eq: torque_mp_relation}
\Lambda \rightarrow \frac{\Lambda}{k^2_{3\textrm{D}/1\textrm{D}}}.
\end{equation}
In other words, for $\alpha=0.01$ we simply divide equation \ref{eq:torqueDensity} by $1.5^2$, and then the pebble isolation masses obtained from equation \ref{eq: mIso B18} are correct even in 1-D. The power of 2 is there because the torque density is proportional to the planetary mass square (Equation \ref{eq:torqueDensity}). 

\begin{table}
\caption{Approximate values of the scalar $k_{3\textrm{D}/1\textrm{D}}$ for discrete values of $\alpha_{\textrm{visc}}$. This scalar tells us that for a specific value of $\alpha_{\textrm{visc}}$, say $0.01$, the planetary mass required to obtain a zero pressure gradient in our 1-D simulations is $1.5$ times lower than the mass obtained using the equation from \citet{Bitsch2018}, which was obtained using 3-D simulations.}
\label{table: bert factor}
\centering
\begin{tabular}{c c}          
\hline\hline
$\alpha_{\textrm{visc}}$ & $k_{3\textrm{D}/1\textrm{D}}$ \\
\hline
    $0.01$ & 1.5 \\
    $0.001$ & 2\\
    $0.0005$ & 2.5 \\
    $0.0001$ & 5 \\
\hline\hline
\end{tabular}
\end{table}

\subsection{Dust evolution}
We adopt the approach of Lagrangian super-particles for the solid component of the disc. Each super-particle represents multiple identical physical solid particles, and each super-particle $i$ has its own position $\vec{x}_i$ and velocity $\vec{v}_i$. The particle velocity is taken as the sum of the drift velocity and the turbulent velocity, the algorithms for which are described below.

\subsection*{Drift velocity}
The radial drift velocity of a dust particle in a disc that is accreting gas is
\begin{equation}
v_{\textrm{D}} =-\frac{2\tau_s}{1 + \tau_s^2}\left(\eta v_{\textrm{kep}} - \frac{1}{2\tau_s}v_{\textrm{R}}\right),
\end{equation}
where $\eta$ is the difference between the azimuthal gas velocity ($v_{\theta}$) and the Keplerian velocity ($v_{\textrm{kep}}$), $v_{\textrm{R}}$ is the gas velocity in the radial direction, and $\tau_s$ is the Stokes number, sometimes referred to as the dimensionless stopping time \citep{Nakagawa1986,Guillot2014}. The $\eta$ parameter is directly related to the pressure gradient of the disc as
\begin{equation}
\eta=-\frac{1}{2}\left(\frac{H}{R}\right)^2\frac{\partial \ln P}{\partial \ln R}.
\end{equation}
The Stokes number for a particle in the Epstein regime is 
\begin{equation}
\tau_s=\frac{s \rho_{\bullet}}{H \rho},
\end{equation}
note that the factor $\sqrt{\pi /8}$ in equation 6 from \citet{Carrera2015} is not included in this work and would not have changed the results significantly. In the above equation $s$ is the particle radius, $\rho$ is the gas density in the mid-plane (related to the gas surface density through $\rho=\Sigma/(2\pi H)$) and $\rho_{\bullet}$ is the solid density. We adopt the value $\rho_{\bullet}=1000\, \textrm{kg}\, \textrm{m}^{-3}$ throughout this work. This density could approximately represent the density of icy pebbles with a significant porosity.

\subsection*{Turbulent velocity}
Solid particles in the protoplanetary disc experience a drag force due to the turbulent motion of gas in the disc. The resultant turbulent diffusion of the particles can be modeled as a damped random walk, which we implement using the algorithm from \citet{Ros2019}. They calculate the turbulent diffusion coefficient ($D$) by applying a force acceleration ($f$) to the particles on a time-scale $\tau_{\textrm{for}}$, and damping the turbulent velocity on the correlation time-scale $\tau_{\textrm{cor}}$. The forcing time-step $\tau_{\textrm{for}}$ is set to equal the time-step of the simulation, and the correlation time is the approximate time-scale over which a particle maintains a coherent direction, calculated as the inverse of the Keplerian angular velocity ($\tau_{\textrm{cor}}=\Omega^{-1}$). As an addition to the algorithm from Ros et al. (2019), the falling of $D$ with Stokes number is implemented here following eq. 37 in \citet{YoudinLithwick2007}. We use a value of 1 for the dimensionless eddy time. 

\subsection*{Particle collisions}
Our model of particle collisions follows the results of \citet{guttler2010}. They combine laboratory collision experiments and theoretical models to show that collisions among dust particles in the disc lead to either sticking, bouncing, or fragmentation. 
The outcome is determined by the mass of the projectile (the lightest of the colliding particles) and the collision velocity, which is calculated as the sum of the relative speed from drift, Brownian motion and turbulent motion. The result of the collision also varies depending on the mass ratio of projectile to target particle and on whether the particles are porous or compact. In our simulations we limit ourselves to porous particles, and draw the line between equal-size particles and non-equal-size particles at a mass ratio of 10 (using effectively only the two upper panels of Figure 11 in \citet{guttler2010}). If the outcome is sticking, then the target mass is either doubled or multiplied by $(1+M_{\textrm{projectile}}/M_{\textrm{target}})$, if the total mass in the projectile particle $M_{\rm projectile}$ is smaller than the total mass in the target particle $M_{\rm target}$. For fragmentation we set all the target and projectile particles to the mass of the projectile. For a complete description of the collision algorithm see Appendix \ref{Appendix: collision algorithm}.

\subsection{Planetesimal formation}\label{subsec: planetesimal formation}
\citet{Carrera2015} performed hydrodynamical simulations of particle-gas interactions to find out under which conditions solid particles in the disc come together in dense filaments that can collapse under self-gravity to form planetesimals. By doing so they mapped out for which solid concentrations and particle Stokes number filaments emerge. This map was revised by \citet{Yang2017} who expanded on the investigation by using longer simulation times and significantly higher resolutions. The critical curves on the map for when the solid concentration is large enough to trigger particle clumping are given by \citet{Yang2017} as
\begin{equation}
\log Z_{\textrm{c}} = 0.3(\log \tau_s)^2 + 0.59\log \tau_s - 1.57 \quad (\tau_s > 0.1), 
\end{equation}
and
\begin{equation}
\log Z_{\textrm{c}} = 0.1(\log \tau_s)^2 + 0.20\log \tau_s - 1.76 \quad (\tau_s < 0.1),
\end{equation}
where $Z_{\textrm{c}}=\Sigma_{\textrm{solid}}/\Sigma_{\textrm{total}}$, and the logarithm is with base 10. These equations were derived for a laminar disc model; however, unless the degree of turbulence is very high they may also be valid for non-laminar discs \citep{Yang2018}. For example, \citet{Yang2018} find a critical solid-to-gas ratio of 2\% when using $\tau_s=0.1$ particles and a vertical turbulence strength of $10^{-3}$, driven by density waves excited by the magnetorotational instability in the turbulent surface layers.

\subsection*{Pressure dependence}

The map from \citet{Yang2017} determines whether or not the streaming instability forms filaments based on the solid abundance and the Stokes number; however, the degree of clumping is also strongly dependent on the radial pressure gradient (\citealt{BaiStone2010_Pgrad,Abod2018,AuffingerLaibe2018}). Simulations by \citet{BaiStone2010_Pgrad} show that the critical solid abundance required to trigger particle clumping via the streaming instability monotonically increases with the radial pressure gradient. In other words, planetesimal formation is most likely to occur inside pressure bumps where the pressure gradient is small, a result which has also been obtained from a linear analysis \citep{AuffingerLaibe2018}.
From \citet{BaiStone2010_Pgrad} it appears that the critical solid abundance is roughly linearly dependent on the radial pressure gradient. In order to catch this dependency we scale the metallicity threshold from \citet{Yang2017} with the local radial pressure gradient divided by the background pressure gradient. Here we made the assumption that the background pressure gradient in our simulations is the same as in \citet{Yang2017}. This is a simplification; however, a comparison of simulations using the different background pressure gradients show that the results are unaffected.

When the pressure gradient is exactly zero there is no particle drift, meaning that the streaming instability is formally absent. However, there is another planetesimal formation mechanism still active which we have not not included in our simulations, namely secular gravitational instability (e.g. \citealt{Youdin2011,TakahashiInutsuka2014}). Recently, \citet{Abod2018} performed simulations of planetesimal formation for various pressure gradient conditions (including a zero pressure gradient). They find that many results and conclusions obtained in their study of planetesimal formation via the streaming instability are valid also in the case of a zero pressure gradient. Motivated by this result we choose to use the same mechanism for planetesimal formation at a pressure gradient of zero as we do otherwise. 

\subsection*{Code implementation}

We have implemented planetesimal formation in the code in the following way: 1) We calculate the solid abundance in each grid cell, excluding the already formed planetesimals; 2) We calculate the local pressure gradient in each grid cell; 3) We scale the metallicity threshold $Z_{\textrm{c}}$ up and down linearly by the found local pressure gradient divided by the background pressure gradient; 4) We calculate the mean Stokes number in each grid cell; 5) If the criterion for planetesimal formation is reached, then we set the radius of the first super-particle in the grid cell to be $100\, {\textrm{km}}$; 6) We repeat the process until the criterion is no longer met. We note that the planetesimal size is arbitrary since we do not follow the dynamical evolution of the planetesimals after their formation.

The algorithm described above implies that every time the conditions for the streaming instability are met, planetesimals form. This can be thought of as a limiting maximum case for the planetesimal formation efficiency, and a discussion regarding how accurate this algorithm is can be found in section \ref{subsec: efficiency of SI}. 

\subsection*{An alternative model for planetesimal formation}

The criterion used in this paper for planetesimal formation via the streaming instability is not the only one that exists. Another commonly used criterion is that the dust-to-gas ratio in the mid-plane has to be larger than unity. Such a planetesimal formation model is used in e.g. \citet{Stammler2019}, who conducts a similar study of planetesimal formation in pressure bumps. We implement this planetesimal formation criterion (which we refer to as the mid-plane model) in the code and make a comparison of the two planetesimal formation models in section \ref{Appendix: compare SI}. For code implementation we use the same algorithm as above, except that it is applied to the mid-plane density ratio rather than to the column density ratio.


\section{Numerical setup}\label{sect: numerical methods}
\bgroup
\def\arraystretch{1.3}%
\begin{table}
\caption{Model set-up for the simulations in the parameter study.}
\label{table: parameter study}
\centering                                      
\begin{tabular}{l l l l l l}          
\hline\hline
Run & $\alpha_{\textrm{visc}}$ & $\alpha_{\textrm{turb}}$ & $\Sigma_{\textrm{solid}}/\Sigma$ & M$_\textrm{ P}$ & V$_\textrm{P}$ \\
 & & & & ($\textrm{M}_{\textrm{iso}}$) & (au/Myr) \\
\hline
    \#1 nominal & $10^{-2}$ & $10^{-3}$ & 0.01 & 2 & 0 \\
    \#2 $0.50\, \textrm{M}_{\textrm{iso}}$ & $10^{-2}$ & $10^{-3}$ & 0.01 & 0.5 & 0 \\
    \#3 $0.75\, \textrm{M}_{\textrm{iso}}$ & $10^{-2}$ & $10^{-3}$ & 0.01 & 0.75 & 0 \\
    \#4 $1\, \textrm{M}_{\textrm{iso}}$ & $10^{-2}$ & $10^{-3}$ & 0.01 & 1 & 0 \\
    \#5 $3\, \textrm{M}_{\textrm{iso}}$ & $10^{-2}$ & $10^{-3}$ & 0.01 & 3 & 0 \\
    \#6 lowVisc & $10^{-3}$ & $10^{-3}$ & 0.01 & 2 & 0 \\
    \#7 lowTurb & $10^{-2}$ & $10^{-4}$ & 0.01 & 2 & 0 \\
    \#8 lowViscTurb & $10^{-4}$ & $10^{-4}$ & 0.01 & 2 & 0 \\
    \#9 highMetal & $10^{-2}$ & $10^{-3}$ & 0.02 & 2 & 0 \\
    \#10 migration & $10^{-2}$ & $10^{-3}$ & 0.01 & 2 & 6.3 \\
\hline\hline
\end{tabular}
\end{table}

The code we use for simulations is called PLANETESYS, and is a modified version of the Pencil Code \citep{BrandenburgDobler2002}, designed for highly parallel calculations of the evolution of gas and dust particles in protoplanetary discs. The code is developed under the ERC Consolidator Grant "PLANETESYS" (PI: Anders Johansen) and this paper together with the recent paper by Ros \& Johansen (2019) represent the first publications using this tool. 

The evolution of the surface density is solved using a first order finite difference scheme with an adaptive time-step. The disc stretches from 1 to $500\, \textrm{au}$ with $R_{\textrm{out}}=100\, \textrm{au}$ and is modelled using a linear grid with 4000 grid cells. For the inner boundary condition we copy the values of the adjacent cells, and for the outer boundary condition we simply set the density to zero at the outer disc edge. This provides the right solution to the viscous disc problem, and fits the analytically derived surface density profile well out to at least $200\, \textrm{au}$. We use a stellar mass of $1\, \textrm{M}_{\odot}$ throughout the simulations and an initial disc accretion rate of $\dot{M}_0=10^{-7}\, \textrm{M}_{\odot}\, \textrm{yr}^{-1}$. The accretion rate drops to $2\times 10^{-8}\, \textrm{M}_{\odot}\, \textrm{yr}^{-1}$ after 1 Myr, as material drains onto the star. Most simulations were run on 40 cores to speed up the calculations. The typical wall time was 90 hours.

Three planets of fixed masses and semimajor axes (except in simulation migration) are included in the simulations, and they are inserted at semimajor axes corresponding to the locations of the major gaps in the disc around the young star HL Tau: at $11.8\, \textrm{au}$, $32.3\, \textrm{au}$ and $82\, \textrm{au}$ \citep{Kanagawa2016}. The solid population of the disc is represented by 100,000 superparticles (approximately 25 per grid cell). The superparticles are initially placed equidistantly throughout the disc with a radius of $1\, \mu m$. The mass of each superparticle is set as to yield a constant solid-to-gas ratio (also referred to as the metallicity) across the disc. Planetesimal formation is initiated after some time $t_{\textrm{plan}}$, which is varied between the simulations depending on how much time it takes for the planets to clear most of their gaps from dust. Further on, we only allow for planetesimal formation interior to $200\, \textrm{au}$. This is because our numerical surface density profile starts to diverge from the analytically derived one beyond a few hundred au; however, since we are only interested in the inner $\sim$100$\, \textrm{au}$ where the planets are located, this does not affect the results. Finally the system is evolved for $1\, \textrm{Myr}$. This long running time is a major motivation for simulating in 1-D only.

In the nominal model (simulation \#1 in Table \ref{table: parameter study}) we use a turbulent viscosity of $10^{-2}$, a turbulent diffusion of $10^{-3}$, and an initial solid-to-gas ratio of $0.01$. We set the planetary masses ($M_{\textrm{p}}$) to be two times their respective pebble isolation mass, and keep the planets at a fixed position ($V_{\textrm{p}}=0$). In this simulation planetesimal formation is initiated after 5,000 years. To explore how the above mentioned parameters affect the planetesimal formation efficiency and the distribution of dust and pebbles, we conduct a parameter study. The parameter values used in the different simulations can be found in Table \ref{table: parameter study}. In simulation \#2-\#5 we vary the planetary masses, in simulation \#6-\#8 we lower the value of the viscosity parameter and turbulent diffusion, in simulation \#9 we increase the initial solid-to-gas ratio in the disc, and in simulation \#10 we let the planets migrate radially inwards with a constant velocity.

\section{Results}\label{sect: results}

Results on the disc structure, particle distribution and efficiency of planetesimal formation in the nominal model are presented in section \ref{subsec: nominal model}. In section \ref{subsec: noPlan_noPscal} we show how these results change when we no longer include the pressure scaling for the streaming instability, and when planetesimal formation is removed completely. In section \ref{subsec: parameter study} we vary different disc and planet parameters and investigate how it affects the results. Finally, in section \ref{Appendix: compare SI} we make a comparison between our model for planetesimal formation and the mid-plane model.

\subsection{Nominal model}\label{subsec: nominal model}

\begin{figure*}[ht]
\resizebox{\hsize}{!}
    {\includegraphics{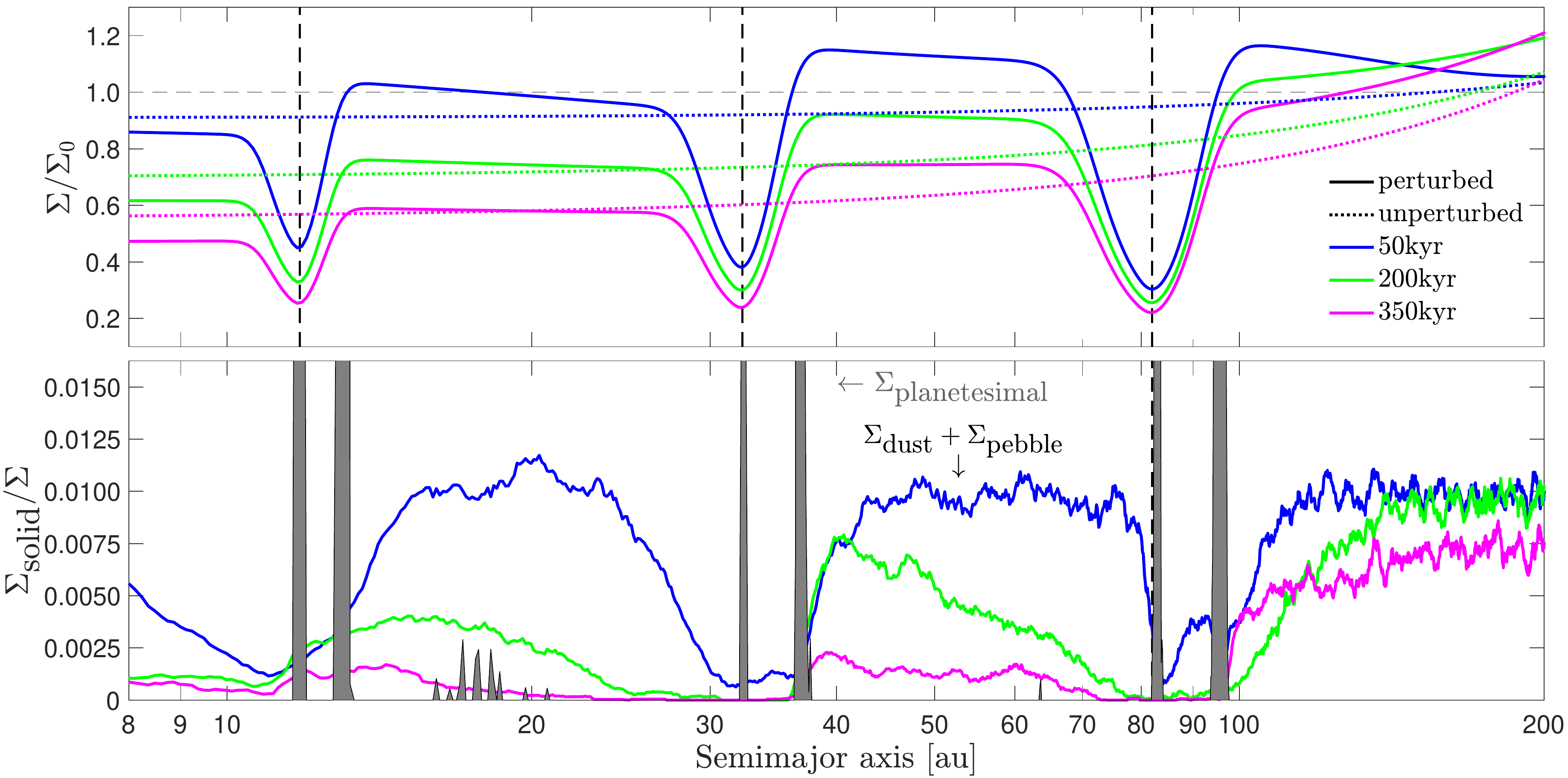}}
\caption{Top panel: Evolution of the gas surface density, normalized against the initial gas surface density, across the protoplanetary disc for the nominal model with planets (solid lines) and without planets (dotted lines). The vertical dashed lines mark the semimajor axes of the planets, and coincide with the locations of the three major gaps in the disc around HL Tau \citep{Kanagawa2016}.
Bottom panel: Evolution of the solid-to-gas surface density ratio across the protoplanetary disc for the nominal model. The solid component is divided into planetesimals (marked with grey), and dust + pebbles. Planetesimals form in narrow rings at the location of the gap edges and inside the planetary gaps (the amount of planetesimals formed inside the gaps is negligible). The interplanetary regions are depleted of dust and pebbles within a few hundred thousand years.}
    \label{fig:sigma_z_nominal}
\end{figure*}
\begin{figure*}
\resizebox{\hsize}{!}
    {\includegraphics{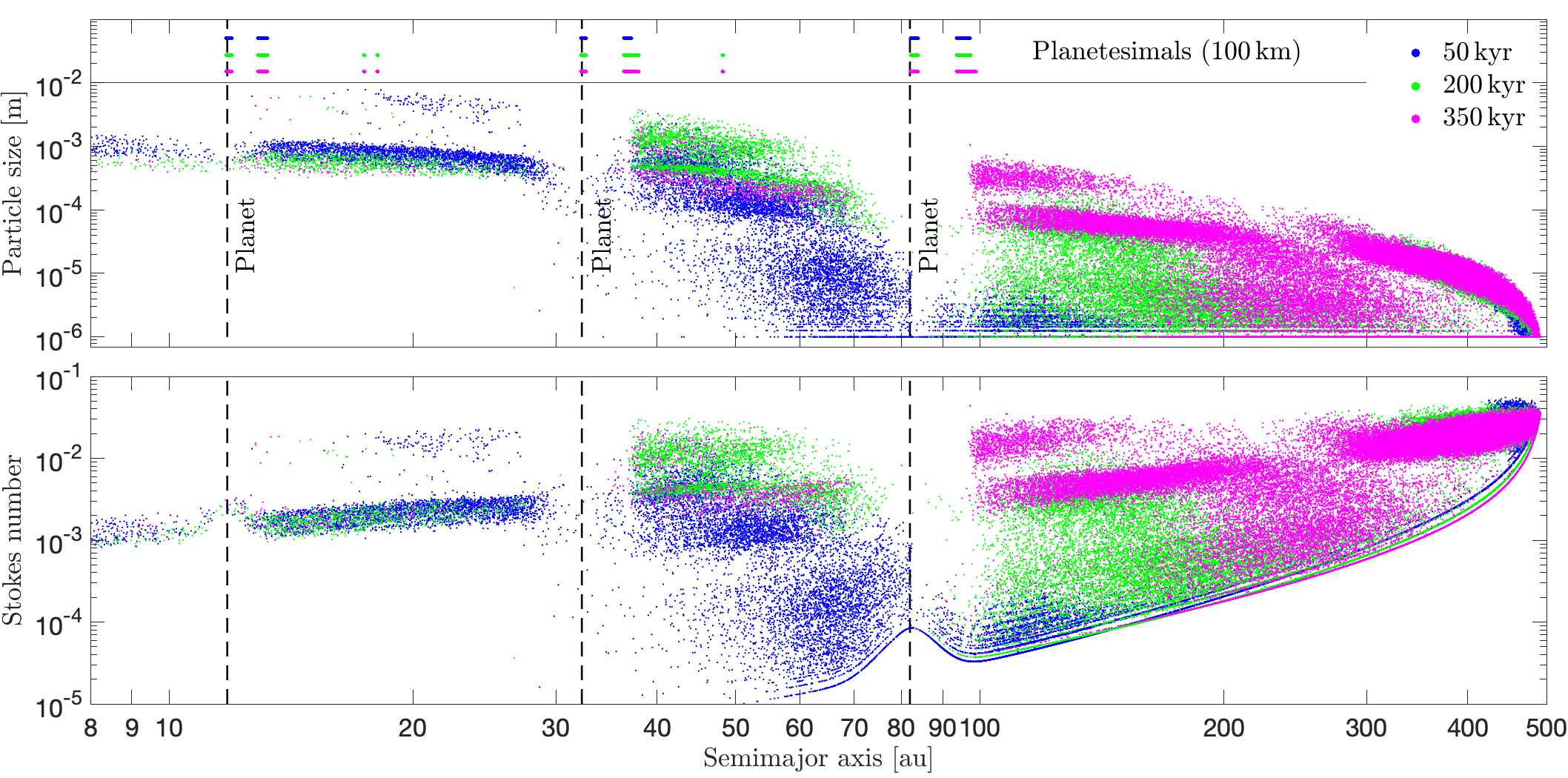}}
\caption{Top panel: Size distribution of particles in the protoplanetary disc at different times during disc evolution for our nominal model. The semimajor axes of the formed planetesimals are indicated at the top of the plot. Particles are initialized with an equal spacing all over the disc and with a radius of $1\, \mu$m. Efficient coagulation and high drift velocities in the inner disc result in a depletion of the interplanetary regions after only a few 100,000 years. These processes occur on a longer timescale in the outer disc. Bottom panel: Plot of particle Stokes number versus semimajor axis for the same data as in the top plot.}
    \label{fig:aVSx_nominal}
\end{figure*}

\begin{figure*}
\centering
\resizebox{0.7\hsize}{!}
    {\includegraphics{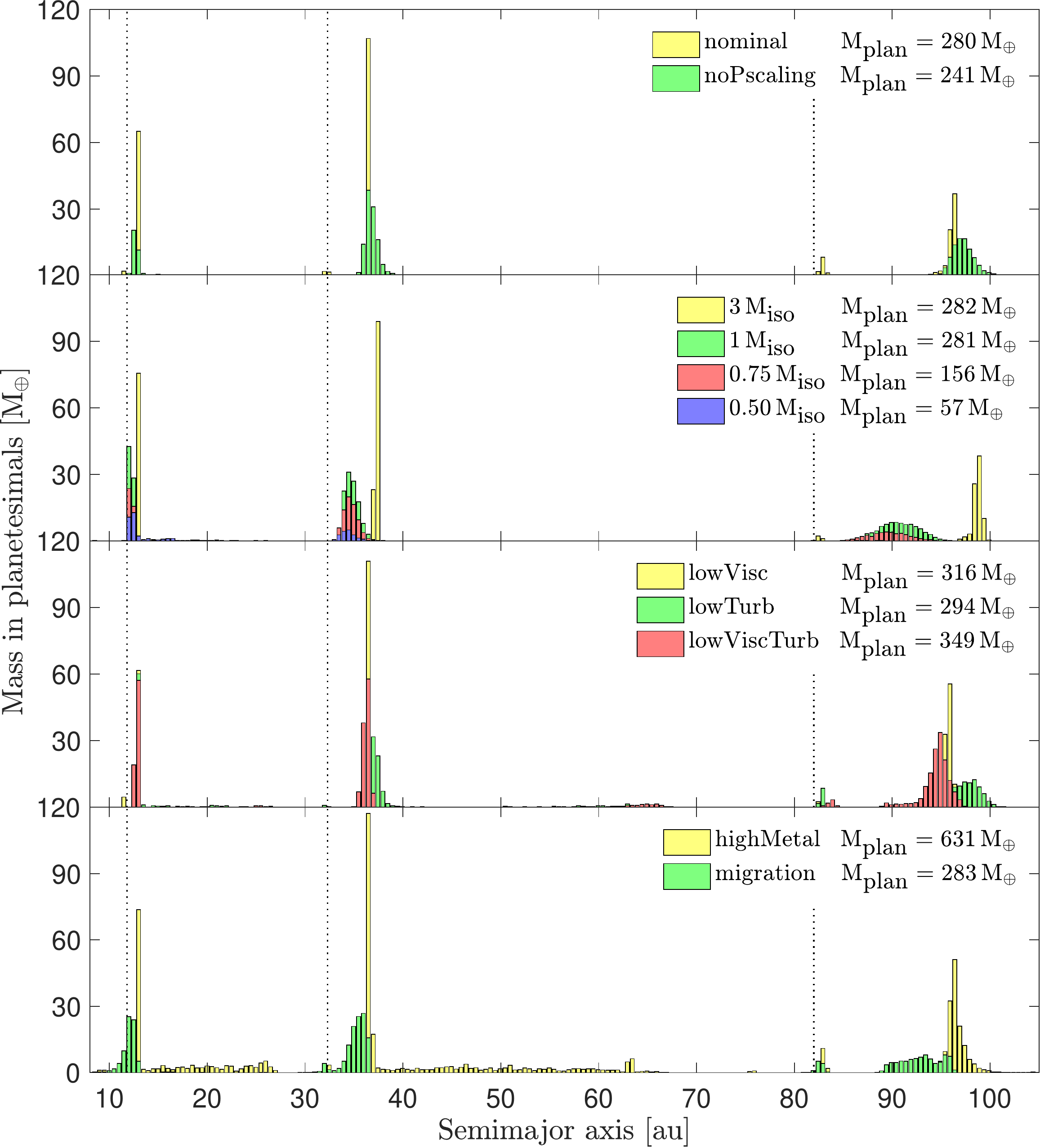}}
\caption{Histogram showing the total amount of mass in planetesimals that has formed at different locations in the disc after $1\, \textrm{Myr}$, for each simulation in the parameter study. The dotted vertical lines indicate the semimajor axes of the planets. When the pressure scaling is removed (simulation ``noPscaling''), the critical density required for forming planetesimals at the gap edge increases, resulting in less planetesimal formation. The amount of planetesimal formation does not vary a lot between the simulations where the planetary mass is larger than the pebble isolation mass, since millimeter pebbles can not drift past the gaps, and planetesimal formation is extremely efficient where the pressure gradient is close to zero. For planetary masses lower than the pebble isolation mass, the amount of planetesimals that form decreases rapidly with decreasing planetary mass. Lowering the viscosity parameter and the turbulent diffusion results in faster drift velocities in the viscously expanding part of the disc, leading to that more pebbles reach the outermost gap edge and are turned into planetesimals. Increasing the metallicity (simulation highMetal) results in sporadic planetesimal formation in the interplanetary regions. When a constant migration speed is added to the planets the location of the gap edges are shifted inwards with time, resulting in that the region where planetesimals form shifts inwards accordingly. }
    \label{fig:hist_all}
\end{figure*}

The parameters used in the nominal model are found in Table \ref{table: parameter study}, simulation \#1. The evolution of the normalized gas surface density across the disc is shown in the top panel of Figure \ref{fig:sigma_z_nominal}, and the evolution of the solid-to-gas surface density ratio is plotted in the bottom panel. The solid component has been divided into planetesimals and dust + pebbles. The evolution of the particle size distribution and the Stokes number is shown in Figure \ref{fig:aVSx_nominal}.

From Figure \ref{fig:sigma_z_nominal} it is clear that the region interior to the outermost planet becomes depleted of dust and pebbles already after a few hundred thousand years. Meanwhile, the region exterior to the outermost planet maintains a high solid-to-gas ratio for much longer, and at the end of the simulation it is still at 25\% of the initial value. To understand why this is we look at the evolution of the particle sizes in Figure \ref{fig:aVSx_nominal}. Considering the inner part of the disc, particle collisions are frequent and quickly result in the formation of mm-sized pebbles which drift fast towards the star. The same process results in smaller particle sizes and takes more time in the outer part of the disc. Another feature of the particle size distribution is that it is bimodal. The reason for this is that except for very low relative velocities, coagulation is only possible if the projectile is much less massive than the target (compare the two upper panels of Figure 11 in \citet{guttler2010}). Hence any coherent size distribution will evolve to a bimodal distribution as the largest particles grow in mass, while the small are stuck. The bimodal size distribution nevertheless collapses with time to a narrower size distribution, as the remaining small particles are finally swept up by the larger pebbles.

The planetary masses used in the nominal simulation lead to relatively deep gaps, which act as hard barriers that stall all particles with a Stokes number larger than a certain critical value (see e.g. \citealt{Pinilla2012,Bitsch2018,Weber2018}). The pebbles formed by coagulation beyond the middle planet have Stokes numbers well above this critical value, and are therefore efficiently trapped at the planetary gap edges. Since no particles make it past the gaps there is no replenishing of the interplanetary disc regions. Combined with fast radial drift, this is the reason why the solid-to-gas ratio in these regions drops to less than 0.001 after only 500,000 years.

From Figure \ref{fig:sigma_z_nominal} and Figure \ref{fig:aVSx_nominal} it is evident that planetesimal formation takes place almost exclusively in narrow rings at the location of the gap edges and inside the planetary gaps. These are locations which correspond to places where the pressure gradient is close to zero, i.e. places where there is a pressure bump. When the pressure gradient is close to zero the critical density required for the streaming instability to form filaments becomes very low, and because of this essentially everything that enters the pressure bump is turned into planetesimals. So since everything that goes into the pressure bump is turned into planetesimals, and there is no replenishing of the interplanetary regions, the region interior to the outermost planet becomes empty of dust and pebbles. As will be seen in section \ref{subsec: noPlan_noPscal}, efficient planetesimal formation at the gap edge of the innermost planet is the reason why no pebbles can make it past that gap to the innermost disc region.

The narrowness of the rings in which planetesimals form is also an effect caused by the dependency of the streaming instability on the pressure gradient. The magnitude and steepness of the pressure bump increases with increasing planetary mass, and for a planetary mass of two pebble isolation masses the pressure gradient is only close to zero in two very narrow regions. In between these regions the pressure gradient becomes very high, resulting in that very large critical densities are required to form filaments, which is why no planetesimals form in between the location of the gap edges and inside the planetary gaps. 

The pressure maxima inside the planetary gaps are unstable equilibrium points due to the fact that tiny displacements from this point will be amplified with time due to divergent drift, but we still form some planetesimals there because we turn on planetesimal formation before the gaps has been entirely cleared of dust and pebbles. Later on during the simulation all planetesimals form at the gap edges. A histogram showing the total amount of planetesimals that has formed per semimajor axis bin can be viewed in the uppermost panel of Figure \ref{fig:hist_all}. A total of 280 Earth masses of planetesimals have formed at the end of the nominal simulation. The total amount of mass in planetesimals and dust + pebbles at the end of all the simulations is provided in Table \ref{table: masses} for each ring. 

\subsection{The cases with no pressure scaling and no planetesimal formation}\label{subsec: noPlan_noPscal}
\begin{figure}
\centering
\includegraphics[width=\hsize]{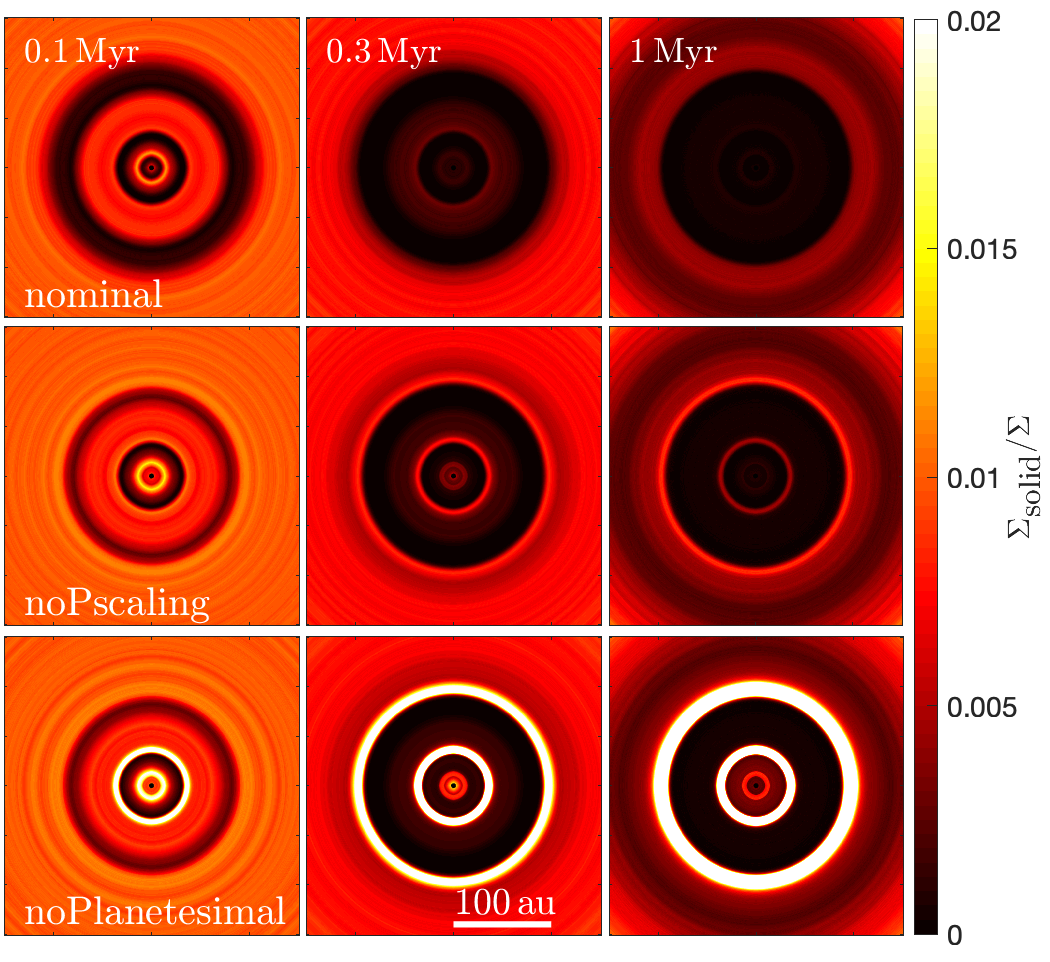}
\caption{2-D symmetric disc images of the evolution of the solid-to-gas surface density (excluding the formed planetesimals) for three different versions of the nominal model: the nominal model (top row), the nominal model with planetesimal formation but no dependency on the pressure gradient (middle row), and the nominal model without planetesimal formation (bottom row). In the nominal model, essentially everything that enters the pressure bump is converted into planetesimals, resulting in a large cavity in the distribution of dust and pebbles. When the pressure dependence is neglected, the critical density required to trigger the streaming instability increases, and we see some rings in the dust and pebble distribution. When planetesimal formation is removed completely, we are left with three rings in which the dust and pebble density is very high. }
\label{fig:discImage_noPlan}
\end{figure}
How efficiently pebbles are converted into planetesimals at the gap edges has a major effect on the distribution of dust and pebbles in the disc. In the nominal model planetesimals form whenever the critical density required for the streaming instability to form filaments has been reached. Together with the linear pressure scaling this can be thought of as a maximum limiting case for planetesimal formation. If the dependency of the streaming instability on the pressure gradient is weakened or removed, the critical density increases, resulting in more dust and pebbles being left in the rings. It would also result in planetesimals forming in wider regions around the pressure bump. The case with no planetesimal formation at all is of course the minimum limiting case for planetesimal formation, and the real efficiency should be somewhere in between the minimum and maximum case. 

In Figure \ref{fig:discImage_noPlan} the evolution of the solid-to-gas surface density ratio is shown as 2-D surface density plots (here referred to as \textit{disc images}) for: the nominal model; the nominal model with no pressure scaling for planetesimal formation; and the nominal model without planetesimal formation. In the case without planetesimal formation (simulation noPlanetesimal) the disc would be seen to have very bright rings at the positions of the outer gap edges, similar to AS 209 \citep{Fedele2018}. The amount of dust + pebbles in each ring is written in Table \ref{table: masses}, and is roughly a hundred Earth masses for the two outermost rings, which corresponds to 3-11 Earth masses per au. In contrast, the inner ring only has a few Earth masses of dust and pebbles. Comparing that to the tens of Earth masses of planetesimals that form in the ring in the nominal model, it suggests that millimeter-sized pebbles with low Stokes numbers (i.e. close to the star) can drift past our planetary gaps quite efficiently, but that the efficient planetesimal formation in the nominal model prevents them from doing so. 

When planetesimal formation without pressure scaling is included (simulation noPscaling) the solid-to-gas surface density ratio drops drastically in the rings. The two rings which are still clearly visible are also much thinner than in the case with no planetesimal formation. The amount of dust and pebbles left in these rings are now on the order of a few Earth masses, which corresponds to a quarter of an Earth mass per au. When pressure scaling is applied (i.e. the nominal model) the amount of early transport through the gaps decreases, resulting in a faster pebble depletion. After $1\, \textrm{Myr}$ all rings previously visible interior to the outermost planet have disappeared, and we are left with a cavity of roughly $100\, \textrm{au}$ in size in the dust and pebble distribution. A comparison between the amount of planetesimals that form, as well as where they form, in simulations with the pressure scaling and without it can be seen in the top panel of Figure \ref{fig:hist_all}. The differences between the simulations in Figure \ref{fig:discImage_noPlan} tells us that two discs that are similar in all aspects except for in the efficiency of planetesimal formation could come across as completely different in observations (see discussion in section 7.3 on the efficiency of the streaming instability).

\subsection{Parameter study}\label{subsec: parameter study}
\begin{figure*}
\centering
\resizebox{.8\hsize}{!}
    {\includegraphics{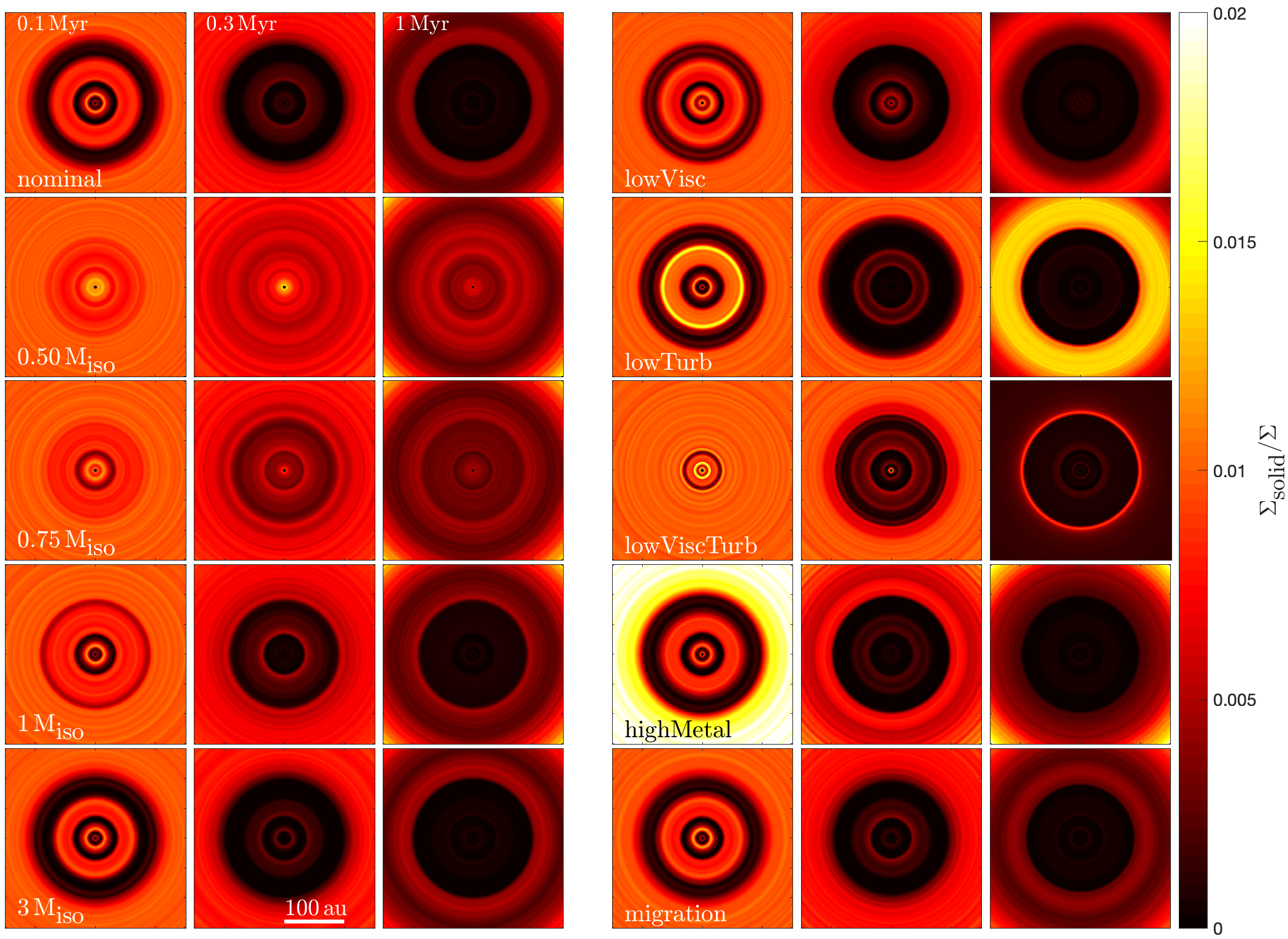}}
\caption{2-D symmetric disc images of the evolution of the solid-to-gas surface density excluding planetesimals for all simulations in the parameter study. In the nominal simulation where we use a planetary mass of 2 times the pebble isolation mass fast pebble drift, little or no transport through the planetary gaps and efficient planetesimal formation at the gap edges result in that the region interior to the outermost planet gets depleted of dust and pebbles. In the simulations where the planetary mass is lower than the pebble isolation mass (simulations $0.50\, \textrm{M}_{\textrm{iso}}$ and $0.75\, \textrm{M}_{\textrm{iso}}$), increased transport past the gaps and less planetesimal formation results in a gap-and-ring like structure. The width of the region where dust and pebbles are trapped depends on the planetary mass, so when the planetary mass is changed to 1 and 3 times the pebble isolation mass (simulations $1\, \textrm{M}_{\textrm{iso}}$ and $3\, \textrm{M}_{\textrm{iso}}$), the width of the rings becomes narrower respectively larger than in the nominal model. Except for this the results are the same as in the nominal model. In simulation lowVisc the viscosity parameter is lowered, resulting in a slower clearing of the gaps and interplanetary regions, and faster radial drift in the viscously expanding part of the disc. When the turbulent diffusion is lowered (simulation lowTurb) the collisional velocities decrease, resulting in larger particles which drift faster towards the star. This causes the bright ring seen beyond the outermost planet at the end of the simulation. In simulation lowViscTurb the turbulent diffusion is kept at the same level as in simulation lowTurb, but the viscosity parameter is lowered by an extra order of magnitude compared to simulation lowVisc. The combination results in that essentially all solids in the outer disc reach the outermost planetary gap before the end of the simulation, causing the narrow bright ring seen in the dust-to-gas ratio. In simulation highMetal the initial solid-to-gas ratio in the disc is increased to 2\%, a change which does not have a big effect on the appearance of the disc at the end of the simulation. Finally in the last simulation the planets were given a constant radial velocity directed towards the star (simulation migration), resulting in a smaller radius of the cavity.}
    \label{fig:discImage_all}
\end{figure*}

To explore how different parameters affect the planetesimal formation efficiency and the distribution of dust and pebbles in the disc, we conduct a parameter study. The values of the parameters which we investigate can be found in Table \ref{table: parameter study}. A histogram showing the total amount of mass locked up in planetesimals at different semimajor axes after $1\, \textrm{Myr}$ is presented in Figure \ref{fig:hist_all} for all simulations in the parameter study. The evolution of the dust and pebble surface density relative to the gas surface density is presented in Figure \ref{fig:discImage_all} as 2-D symmetric disc images for the same simulations. The total amount of mass in planetesimals and dust + pebbles in each ring can be found in Table \ref{table: masses}. The evolution of the particle size distribution for a few interesting simulations in the parameter study is shown in Figure \ref{Appendix: fig size dist} of Appendix. 

\subsection*{Planetary mass}
The planetary mass is the main controller of the width and depth of the planetary gap, as well as the radial pressure gradient. The width of the gap determines where pebbles are trapped, and thus the location of planetesimal formation. The planetesimal formation efficiency is strongly related to the strength of the pressure maxima, both via the scaling of the streaming instability with the radial pressure gradient and via the efficiency of particle trapping. The distribution of dust and pebbles for simulations with varying planetary masses is shown in Figure \ref{fig:discImage_all} (simulations nominal, $0.50\, \textrm{M}_{\textrm{iso}}$, $0.75\, \textrm{M}_{\textrm{iso}}$, $1\, \textrm{M}_{\textrm{iso}}$ and $3\, \textrm{M}_{\textrm{iso}}$).

For the simulations $0.50\, \textrm{M}_{\textrm{iso}}$ and $0.75\, \textrm{M}_{\textrm{iso}}$ we used a planetary mass lower than the pebble isolation mass. In these simulations dust and pebbles are partly transported through the planetary gaps, resulting in a continuous replenishing of the interplanetary regions and the region interior to the innermost planet. There is a small pile-up of pebbles at the gap edges, resulting in a gap-and-ring like structure. As can be seen in Figure \ref{fig:hist_all}, the amount of planetesimals that form decreases fast with decreasing planetary mass. By lowering the planetary mass to 25\% below the pebble isolation mass, the amount of pebbles converted into planetesimals is halved. Since the radial pressure gradient is shallow around the gap edges, planetesimals form in relatively wide regions, and not in two narrow rings like in the nominal model. A plot of the size distribution of particles for simulation $0.75\, \textrm{M}_{\textrm{iso}}$ can be viewed in Figure \ref{Appendix: fig size dist} of the Appendix.

In the simulations nominal, $1\, \textrm{M}_{\textrm{iso}}$ and $3\, \textrm{M}_{\textrm{iso}}$ the planetary masses are 2, 1 and 3 times the pebble isolation mass. For these cases the amount of mass converted into planetesimals does not change with increasing planetary mass. This is due to two reasons: 1) the planetary gaps act as hard barriers which prevents any pebbles from passing through; 2) there are pressure maxima outside all gaps which efficiently turns most or all pebbles into planetesimals. The locations where planetesimals form nevertheless do change a bit, since the widened gap and the steepened pressure gradient result in planetesimals forming further away from the planet and in narrower regions. In summary, all simulations with planetary masses equal to or above the pebble isolation mass appear as discs with large central cavities, with the only difference being a slight dependence of the width of the cavity and the width of the rings on the planetary mass.

\subsection*{Viscous and turbulent $\alpha$}
The viscosity parameter $\alpha_{\textrm{visc}}$ governs the gas accretion rate onto the central star, and it also enters the particle drift equation via the radial gas velocity $v_{\textrm{R}}$. The turbulent diffusion $\alpha_{\textrm{turb}}$ governs the turbulent speed of particles, which in turn affects the frequency of particle collisions as well as the collision velocities. The values inferred from observations of these parameters vary a lot in the literature. In \citet{Pinte2016} they find a value of a few times $10^{-4}$ for the turbulent diffusion in the disc around HL Tau. They obtained this value by assuming a standard dust settling model, varying the amount of turbulent diffusion, and comparing the resulting millimeter dust scale heights to observations. \citet{Pinte2016} further report an upper limit to the viscosity parameter of $10^{-2}$ for the HL Tau disc, a value which was calculated using an estimate of the disc accretion rate from \citet{Beck2010}. Other examples are \citet{Flaherty2017} who reported an upper limit of $0.003$ for the turbulent diffusion in HD 163296, and \citet{Flaherty2018} who reported an upper limit of $0.007$ for the viscosity parameter in TW Hya. 

In the nominal model we use a value of $10^{-2}$ for the viscosity parameter, and $10^{-3}$ for the turbulent diffusion. In simulation lowVisc the viscosity parameter was lowered from $10^{-2}$ to $10^{-3}$. In the simulation lowTurb the turbulent diffusion was decreased by a factor of 10 from $10^{-3}$ to $10^{-4}$. Finally in simulation lowViscTurb the viscosity parameter was decreased by another order of magnitude to $10^{-4}$, while the turbulent diffusion was kept at the same level as in simulation lowTurb. When lowering the viscosity parameter the initial disc accretion rate is reduced accordingly, ensuring the same initial disc mass.   

Lowering the viscosity parameter in simulation lowVisc results in lower gas accretion rates onto the star, and thus the gas disc evolves on much longer time-scales. Because of this the disc does not expand as much, resulting in a smaller disc size. Another effect on the structure of the gas disc is that there is a more pronounced pile-up of gas at the inner and outer edges of the planetary gaps. Looking at Figure \ref{fig:discImage_all} and comparing simulation lowVisc to the nominal simulation we see that lowering the viscosity parameter results in a slower clearing of the gaps and interplanetary regions. We also find that pebbles in the viscously expanding part of the disc drift inwards faster. A smaller viscosity parameter results in the velocity component directed outwards becoming lower, making it easier for particles to drift inwards. The faster drift velocities results in more pebbles reaching the outermost planetary gap edge and being turned into planetesimals. This is the reason why slightly more planetesimals are formed in this simulation compared to the nominal one. 

In simulation lowTurb the lowering of the turbulent diffusion results in fewer collisions. Because the particle Stokes numbers in our simulations are low, it also results in lower collisional velocities (if the particle Stokes numbers would have been larger so that the particles start to sediment towards the mid-plane, this would not have been the case, as the decrease in turbulent velocity would have been compensated for by a decrease in dust scale height). Fewer collisions lead to slower particle growth; however, slower collisional velocities result in a larger maximum pebble size. These larger particles obtain higher drift velocities, resulting in many more pebbles reaching the gap edge of the outermost planet. This is the reason why we see a wide and bright ring in the solid-to-gas surface density at $1\, \textrm{Myr}$ for simulation lowTurb in Figure \ref{fig:discImage_all}. The larger particle sizes also require a smaller critical density to trigger the streaming instability, allowing for planetesimals to form further away from the pressure maxima (see Figure \ref{fig:hist_all}). 

In simulation lowViscTurb fast particle drift towards the star in the outer disc now results in all pebbles still remaining at the end of the simulation being concentrated in a narrow bright ring just beyond the outermost planet. Since more pebbles reach the outermost planetary gap edge, the amount of planetesimal formation at that location has increased compared to the amount in the simulations lowVisc and lowTurb. The gas pile-up at the gap edges is also more pronounced, causing some particles to become trapped at the inner gap edges and leading to some planetesimal formation there (see Figure \ref{Appendix: fig size dist} in Appendix for a plot of the particle size distribution for simulations lowTurb and lowViscTurb). 

\subsection*{Metallicity}
In simulation highMetal the initial solid-to-gas ratio was doubled, naturally resulting in the total amount of formed planetesimals increasing. Since the initial solid abundance is closer to the critical value required for the streaming instability to operate, random local concentrations of pebbles now result in some planetesimal formation in the interplanetary regions. However, since the amplitude of the fluctuations caused by diffusion would be much smaller if a physical number of particles were used, this effect is purely numerical.

\subsection*{Planet migration}
In the final simulation of the parameter study (simulation migration) all three planets were given a constant velocity directed towards the star. The migration speed was set to be $6.3\, \textrm{AU}\,\textrm{Myr}^{-1}$, and thus the semimajor axes of the planets after $1\, \textrm{Myr}$ are $5.48\, \textrm{au}$, $25.98\, \textrm{au}$ and $75.676\, \textrm{au}$. This migration speed is not large enough to significantly perturb the shape of the gap. For faster migrating planets hydrodynamical studies have shown that the impact on the structure of the disc can be big (see e.g. \citealt{Li2009,Meru2019,Nazari2019}). 

Since the migration speed in our simulation is both low enough to preserve the shape of the gap, and significantly lower than the particle drift velocity, the amount of planetesimal formation does not change. The location where they form do however change. As the pressure maximum moves inwards so does the region of planetesimal formation. For the pebble distribution the only thing that changes is that the radius of the cavity shrinks with time.

\subsection{Two planetesimal formation models}\label{Appendix: compare SI}
\begin{figure}
\centering
\includegraphics[width=\hsize]{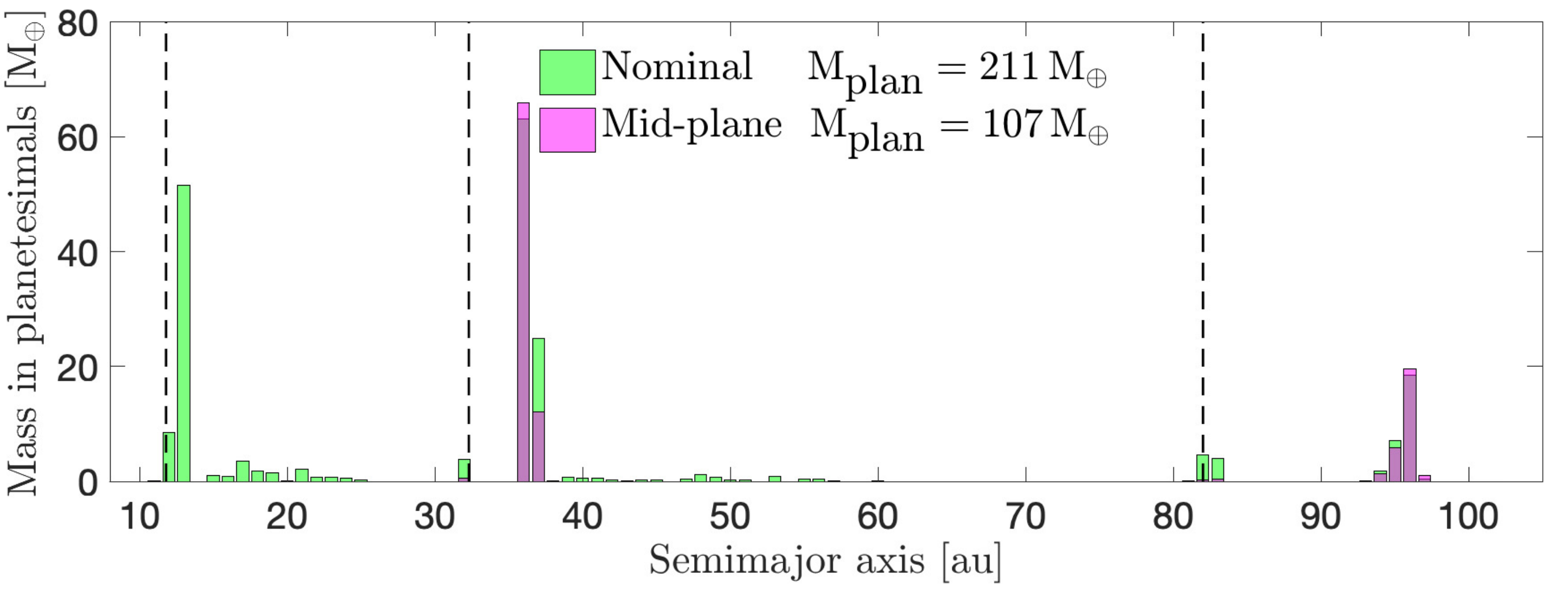}
\caption{Histogram showing the total amount of mass in planetesimals that have formed at different locations in the disc after $300\, \textrm{kyr}$, for the nominal model and the mid-plane model. The amount of planetesimals formed around the gap edges of the two outermost planets are similar in both models; however, this is not the case at the innermost planetary gap edge. The settling towards the mid-plane is not efficient enough to counteract the stirring by turbulence in this region, and thus a dust-to-gas density ratio of unity in the mid-plane is never reached. }
\label{Appendix: fig compare SI hist}
\end{figure}

\begin{figure}
\centering
\includegraphics[width=\hsize]{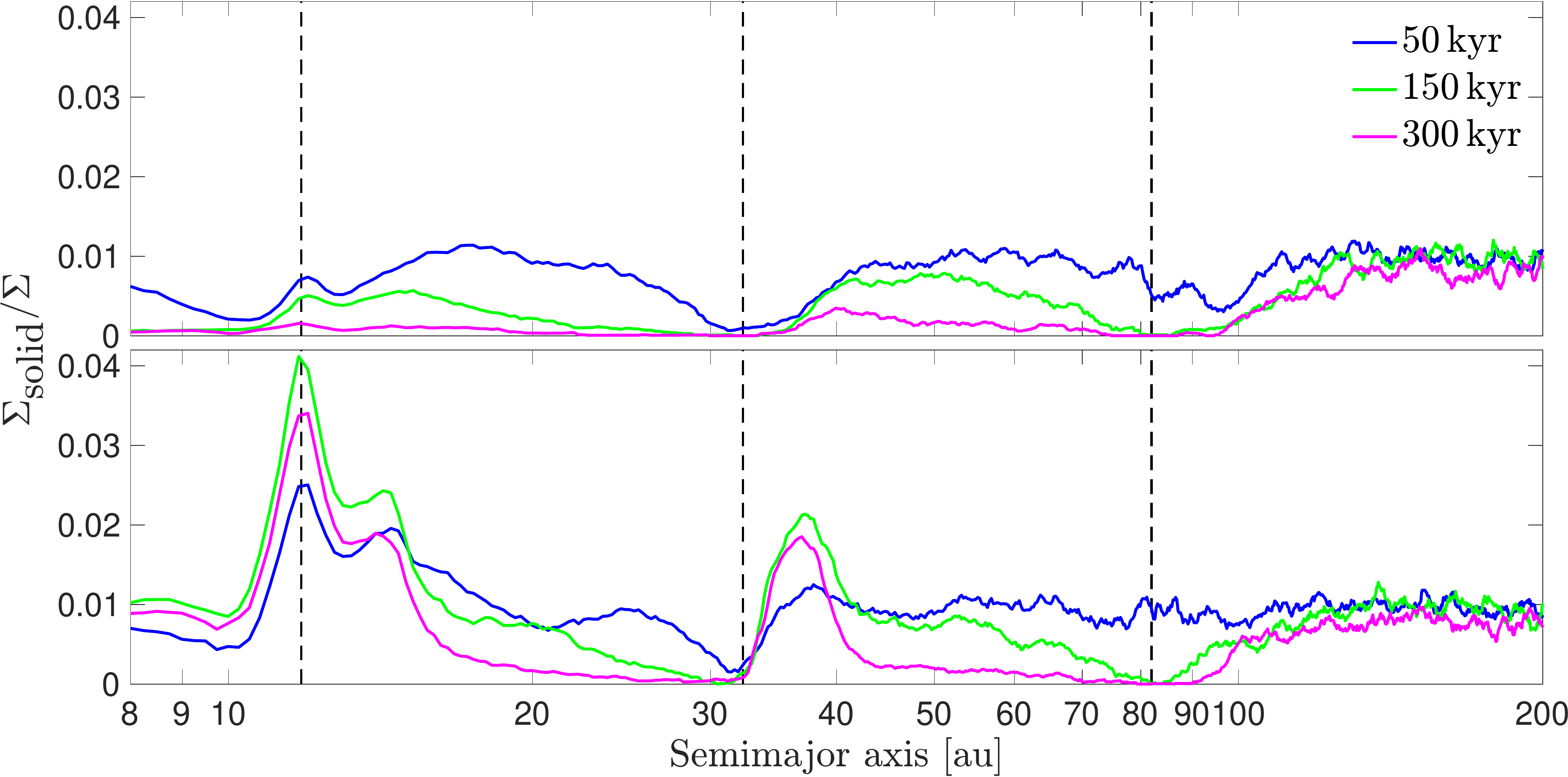}
\caption{Evolution of the solid-to-gas surface density ratio across the protoplanetary disc for the nominal model (top panel) and the mid-plane model (bottom panel). The solid component includes dust and pebbles, planetesimals are excluded. The two models produce very similar amounts of planetesimals around the outermost planetary gap-edge, resulting in very similar solid-to-gas surface density ratios in this part of the disc. Around the second planetary gap-edge the mid-plane model is not quite as efficient at forming planetesimals as the nominal model, and leaves behind a ring of dust and pebbles that does not exist in the nominal model. At the innermost planetary gap-edge there is no planetesimal formation at all in the mid-plane model, and thus the solid-to-gas surface density ratios in the two models are very different. }
\label{Appendix: fig compare SI dustToGas}
\end{figure}

The planetesimal formation model used in our simulations derives from hydrodynamical simulations of particle-gas interactions by \citet{Carrera2015} and \citet{Yang2017}. This model tells us whether or not filaments emerge based on the combination of particle Stokes number and surface density. As mentioned at the end of section \ref{subsec: planetesimal formation}, some authors use a criterion based on the mid-plane dust-to-gas ratio instead (the mid-plane model). We compare this criterion with the one used in this paper by performing two simulations that are identical in all aspects except for the planetesimal formation model. We use a linear pressure scaling in both simulations. The results of the comparison are presented in Figure \ref{Appendix: fig compare SI hist} and Figure \ref{Appendix: fig compare SI dustToGas}. 

The two models produce very similar amounts of planetesimals at the outermost planetary gap-edge, with no significant differences in the surface density profiles. Around the second planetary gap-edge the mid-plane model produces slightly less planetesimals than the nominal model, resulting in a ring of pebbles that does not exist in the nominal model. At the innermost planetary gap-edge the mid-plane model fails at producing any planetesimals. This is because the settling towards the mid-plane is not efficient enough to counteract the stirring by turbulence. Because of this a large population of dust and pebbles remain at this location. Transport of solids through the planetary gap further results in that the innermost disc region does not get depleted of solids within $300\, \textrm{kyr}$, which is the case in the nominal model. 

In the simulations above we have taken into account the dependency of the streaming instability with the pressure gradient. If the linear pressure scaling where to be removed, the mid-plane model does not produce any planetesimals at all. This is because the millimeter-sized pebbles created through coagulation are stirred too much by turbulence, and thus a mid-plane dust-to-gas ratio of unity is never reached. 

\section{Lowering the particle size to $100\, \mu \textrm{m}$}\label{sect:100micron}

\begin{figure}
\centering
\includegraphics[width=.8\hsize]{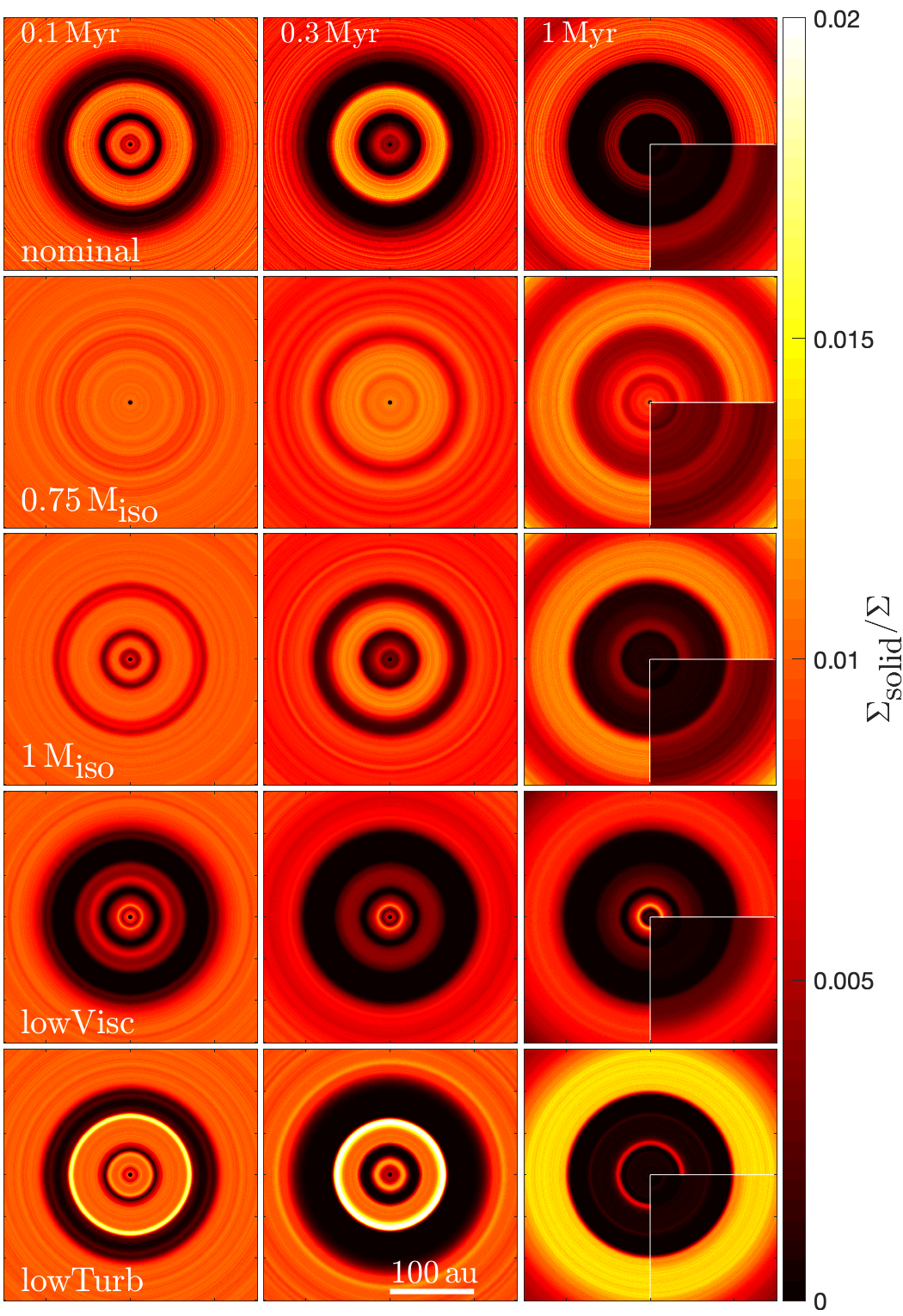}
\caption{2-D symmetric disc images for simulations where the maximum particle size was limited to $100\, \mu \textrm{m}$. For the sake of easy comparison images of the same simulations but without the maximum limit on the particle size are added on top of the images at $1\, \textrm{Myr}$. With smaller particle sizes it becomes harder to trigger planetesimal formation; the result is that more dust and pebbles remain in the rings at the end of the simulation.}
\label{fig:discImage_all_100micro}
\end{figure}

The dust growth model of \citet{guttler2010} employed in this work, which is based on a combination of laboratory collision experiments and theoretical models, results in the formation of mm-sized pebbles in the inner part of the disc, with decreasing grain sizes as the semimajor axis increases. These sizes are slightly smaller than the grain size estimates obtained from the spectral index of the dust opacity coefficient at millimeter and submillimeter wavelengths, which report maximum grain sizes between 1 millimeter in the outer disc and a few centimeter in the inner disc (e.g. \citealt{Ricci2010,Perez2012,ALMA2015,Tazzari2016}). The estimates based on the spectral index of the dust opacity coefficient nevertheless do not agree with the maximum grain sizes obtained from observations of polarized emission due to self-scattering, which are consistently around $100\, \mu \textrm{m}$, i.e. smaller than the pebbles in our simulations (e.g. \citealt{Kataoka2017,Hull2018,Ohashi2018,Mori2019}).

Even when applied to the same source, the maximum grain sizes obtained from the different methods are inconsistent. Consider the disc around HL Tau. \citet{CarrascoGonzalez2019} calculated the maximum grain size in HL Tau by fitting the millimeter spectral energy distribution without any assumptions about the optical depth of the emission. By also including the effects of scattering and absorption in the dust opacity, they obtained a maximum grain size of a few millimeter. \citet{Kataoka2017} instead estimated the maximum grain size in HL Tau to be $100\, \mu \textrm{m}$ from observations of millimeter-wave polarization.

If the maximum grain size was in fact around $100\, \mu \textrm{m}$, as suggested by observations of millimeter-wave polarization, it could have a large impact on our results. In the model of \citet{guttler2010} particle collisions result in the formation of mm-sized particles; however, recently \citet{OkuzumiTazaki2019} showed that dust growth models can result in $100\, \mu \textrm{m}$-sized particles if the particles are covered by nonsticky CO$_2$ ice. By incorporating the composition-dependent sticking into a model of dust evolution they were able to successfully reproduce the polarization pattern seen in the disc around HL Tau. 

Decreasing the particle size results in smaller particle Stokes numbers; however, it should be mentioned that this is not the only way to obtain low particle Stokes numbers. If gas discs are in fact much more massive than the minimum mass solar nebula, then millimeter-sized particles would have smaller Stokes numbers than they have in our simulations \citep{Powell2019}. In such a disc particle drift would be slower, and the critical density required for the streaming instability to form filaments would be higher.

\subsection{Imposing a maximum particle size of $100\, \mu \textrm{m}$ in simulations}\label{subsect:100micron simulations}

We study the effect of having a maximum grain size of $100\, \mu \textrm{m}$ in our simulations by artificially imposing the maximum pebble size to be $100\, \mu \textrm{m}$ in the coagulation part of our code. In the bottom panel of Figure \ref{Appendix: fig size dist} in Appendix \ref{Appendix: sec size dist} we show the resulting particle size distribution across the disc for the nominal simulation. The solid-to-gas ratios for the following simulations: nominal, $0.75\, \textrm{M}_{\textrm{iso}}$, $1\, \textrm{M}_{\textrm{iso}}$, lowVisc and lowTurb, with 100 $\mu$m particles are shown in Figure \ref{fig:discImage_all_100micro}. 

In the nominal simulation, $218\, \textrm{M}_{\oplus}$ of planetesimals are formed. The amount of dust and pebbles left in the rings is larger compared to the nominal simulation with millimeter-sized particles (see Table \ref{table: masses}). Regarding the distribution of pebbles, the Stokes number at the location of the outermost planet is still large enough to prevent most pebbles from drifting across the gap. At the locations of the two innermost planets this is no longer true. However, because of the linear scaling of the streaming instability with the pressure gradient, pebbles at the gap edges are turned into planetesimals before they have time to drift across the gaps. Therefore, the region interior to the middle planet becomes depleted of dust and pebbles. 

When the planetary mass is decreased to $0.75\, \textrm{M}_{\textrm{iso}}$, the global solid-to-gas ratio remains at around 1\% throughout the simulation (see second row of Figure \ref{fig:discImage_all_100micro}) . There are still some rings and gaps visible in the particle distribution, and the amount of planetesimals formed is now $80\, \textrm{M}_{\oplus}$. For a planetary mass of $1\, \textrm{M}_{\textrm{iso}}$ it takes more time to clear the interplanetary regions of pebbles than in the nominal simulation, but at the end of the simulation the solid-to-gas density ratio across the disc looks very similar. The total mass in planetesimal is $232\, \textrm{M}_{\oplus}$ for this simulation. 

A lower viscosity (simulation lowVisc) results in faster drift in the viscously expanding part of the disc and more planetesimal formation at the outer gap edge -- in total $237\, \textrm{M}_{\oplus}$ of planetesimals are formed in this simulation. There is no planetesimal formation at the innermost gap edge in this simulation, and instead there is 10 Earth masses of pebbles trapped in that ring, corresponding to around 5 Earth masses per au. The amount of pebbles left in the ring outside the middle planet has also increased compared to the nominal model. 

When the turbulent diffusion was lowered by an order of magnitude to $10^{-4}$ (simulation lowTurb), the amount of planetesimals formed decreased to $192\, \textrm{M}_{\oplus}$. Lower collisional velocities result in particles growing to $100\, \mu \textrm{m}$ further out in the disc. These particles obtain higher drift velocities and reach the gap edge of the outermost planet within a million years, causing the wide and bright ring seen in the bottom row of Figure \ref{fig:discImage_all_100micro}. 

\begin{figure}
\centering
\includegraphics[width=\hsize]{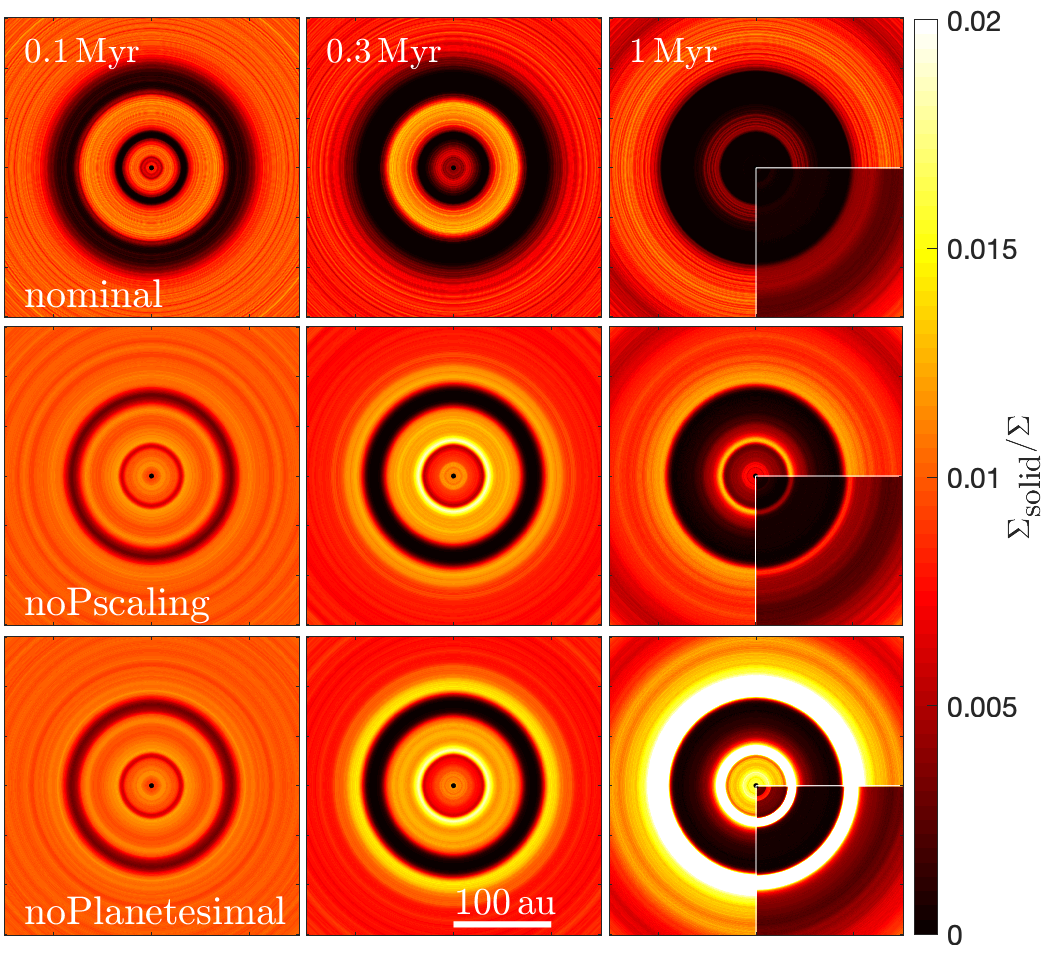}
\caption{2-D symmetric disc images of the evolution of the solid-to-gas surface density for three different versions of the nominal model where the maximum grain size was limited to $100\, \mu \textrm{m}$: the nominal model (top row), the nominal model with planetesimal formation but no dependency on the pressure gradient (middle row), and the nominal model without planetesimal formation (bottom row). For the sake of easy comparison images of the same simulations but without the maximum limit on the particle size are added on top of the images at $1\, \textrm{Myr}$. When the dependency on the pressure gradient is included, efficient planetesimal formation at the gap edges prevents particles from passing through the gaps, resulting in a depletion of the region interior to the middle planet. This does not happen in the other cases, instead several bright rings are seen in the dust-to-gas ratio, and there is dust left also in the innermost region of the disc. }
\label{fig:discImage_noPlan_100micro}
\end{figure}

\subsection{The cases with no pressure scaling and no planetesimal formation}\label{subsec: noPlan_noPscal_100micron}

Next we study how the distribution of dust and pebbles change when 1) the dependency of the streaming instability on the pressure gradient is removed, and 2) when planetesimal formation is removed completely (analogous to Section \ref{subsec: noPlan_noPscal} but for the case with a maximum grain size of 100 $\mu$m). The results are presented in Figure \ref{fig:discImage_noPlan_100micro}. 

There is little or no transport through the outermost planetary gap in all simulations; however, the  100 $\mu$m sized pebbles do drift past the two innermost gaps in the simulation without planetesimal formation (simulation noPlanetesimal). In the simulation without pressure scaling (simulation noPscaling) some of the pebbles that would otherwise have made it past the gaps are now converted into planetesimals instead, resulting in a quicker depletion of the region interior to the middle planet. Apart from this simulation noPlanetesimal and simulation noPscaling result in relatively similar images, the major difference being the brightness and width of the rings. When pressure scaling is added (simulation nominal) the picture changes quite a bit. Efficient planetesimal formation now prevents most pebbles from crossing the middle planetary gap, resulting in that the region interior to this becomes void of pebbles. 

\begin{table*}
\caption{Table containing the total amount of mass in dust/pebbles and planetesimals in each ring at the end of the simulations. The inner ring edges are chosen to be the semimajor axes of the planets, and the outer ring edges are chosen to be four gas scale heights away from this location. The region where particles pile-up varies with e.g. the planetary mass and the level of turbulence in the disc, but for simplicity and for the sake of easy comparison we use the same criteria for the ring widths in all simulations. The gas scale height at the location of the three planets are starting at the inner one: $0.58\, \textrm{au}$, $2.12\, \textrm{au}$ and $7.03\, \textrm{au}$. These are the gas scale heights at the initial locations of the planets. In simulation \#10 the final semimajor axes of the planets are $5.48\, \textrm{au}$, $25.98\, \textrm{au}$ and $75.676\, \textrm{au}$. The corresponding gas scale heights are: $0.22\, \textrm{au}$, $1.61\, \textrm{au}$ and $6.34\, \textrm{au}$.}
\label{table: masses}
\centering
\begin{tabular}{lllllll}
\hline\hline
                         & \multicolumn{2}{c}{Inner ring} & \multicolumn{2}{c}{Middle ring} & \multicolumn{2}{c}{Outer ring} \\ \hline
Run                      & dust/pebbles  & planetesimals  & dust/pebbles   & planetesimals  & dust/pebbles  & planetesimals  \\ \hline
\textbf{\#1 nominal}     & $0.01\, \textrm{M}_{\oplus}$         & $68\, \textrm{M}_{\oplus}$        & $0.1\, \textrm{M}_{\oplus}$          & $116\, \textrm{M}_{\oplus}$        & $3.7\, \textrm{M}_{\oplus}$         & $94\, \textrm{M}_{\oplus}$         \\
\textbf{noPscaling}      & 0.05         & 34         & 2.1          & 110        & 7.7         & 90         \\
\textbf{noPlanetesimal}  & 2.7         &                & 95         &                & 97        &                \\
\textbf{\#2 0.50$\, \textrm{M}_{\textrm{iso}}$}     & 1.3         & 27         & 2.7          & 19         & 8.3         & 2.1          \\
\textbf{\#3 0.75$\, \textrm{M}_{\textrm{iso}}$}    & 1.0         & 40         & 2.4          & 72         & 6.4         & 41         \\
\textbf{\#4 1$\, \textrm{M}_{\textrm{iso}}$}       & 0.007         & 71         & 0.2          & 116        & 4.7         & 92         \\
\textbf{\#5 3$\, \textrm{M}_{\textrm{iso}}$}       & 0.1         & 76         & 0.1          & 122        & 3.5         & 84         \\
\textbf{\#6 lowVisc}     & 0.5         & 69         & 0.1          & 120        & 3.7         & 126        \\
\textbf{\#7 lowTurb}     & 0.02         & 66         & 0.07          & 105        & 11        & 96         \\
\textbf{\#8 lowViscTurb} & 0.6         & 76         & 1.8          & 109        & 13        & 151        \\
\textbf{\#9 highMetal}   & 0         & 83         & 0.03          & 152        & 1.7         & 169        \\
\textbf{\#10 migration}  & 0         & 0          & 0.002          & 2.0          & 2.7         & 94         \\ \hline
\multicolumn{7}{c}{Maximum grain size $100\, \mu m$}                                                                             \\ \hline
\textbf{\#1 nominal}     & 0.05         & 39         & 1.3          & 112        & 5.8         & 65         \\
\textbf{noPscaling}      & 1.8         & 0.02          & 7.5          & 55         & 16        & 48         \\
\textbf{noPlanetesimal}  & 4.0         &                & 28         &                & 61        &                \\
\textbf{\#3 0.75$\, \textrm{M}_{\textrm{iso}}$}    & 2.2         & 0.1          & 6.0          & 37         & 15        & 38         \\
\textbf{\#4 1$\, \textrm{M}_{\textrm{iso}}$}       & 0.1         & 55         & 1.7          & 106        & 12        & 66         \\
\textbf{\#6 lowVisc}     & 10        & 0.02          & 4.6          & 60         & 3.2         & 106        \\
\textbf{\#7 lowTurb}     & 0         & 40         & 3.3          & 95         & 12        & 71         \\ 
\hline\hline
\end{tabular}
\end{table*}

\section{Comparison to observations}\label{sec: observations}
\subsection{Dust mass estimates in rings} \label{subsec: dust mass rings}
The amount of dust and pebbles remaining in the rings after $1\, \textrm{Myr}$ varies a lot in our simulations (see Table \ref{table: masses}). As an example, the amount of dust and pebbles remaining in the outermost ring ranges from $1.7$ to $13$ Earth masses for simulations in the parameter study. We compare these amounts to dust mass estimates by \citet{Dullemond2018} for rings in the DSHARP survey.


\citet{Dullemond2018} found that the amount of dust stored in each ring is of the order tens of Earth masses. As an example, AS 209 was estimated to have around 30 Earth masses of dust trapped in the ring at $69-79\, \textrm{au}$, and 70 Earth masses trapped in the ring at $115-125\, \textrm{au}$. In general the amount of dust stored in the rings ranges from 1 to 10 Earth masses per au \citep[see Table 2 in][]{Dullemond2018}. It should be mentioned that there is much uncertainty in these estimates, mainly due to the uncertainty in the calculation of the dust opacities.  

Comparing with our simulations, the only cases where we have more than 10 Earth masses of pebbles remaining in one or several narrow rings after $1\, \textrm{Myr}$ are when we either 1) ignore planetesimal formation completely; 2) ignore the pressure scaling; 3) use a maximum pebble size of $100\, \mu \textrm{m}$ (note: the mid-plane model is discussed separately below). From table \ref{table: masses} we find that we have between 2-11 Earth masses of dust and pebbles per au left in the rings when planetesimal formation is neglected (simulation noPlanetesimal). When the pressure scaling is removed (simulation noPscaling) this value decreases to 0-0.65 Earth masses per au. Another example is simulation lowVisc for $100\, \mu \textrm{m}$-sized particles, where we find 5 Earth masses of dust and pebbles per au left in the inner ring, and 0.5 Earth masses per au left in the middle ring.

In order to match the dust mass estimations in the DSHARP rings we would thus either need a very low planetesimal formation efficiency, or some mechanism for destroying the planetesimals and thus replenishing the dust population in the rings (this is discussed in Section \ref{subsec:destruction planetesimals}). One mechanism which would result in a higher dust population and likely lead to less planetesimal formation is efficient fragmentation in the pressure bumps; see Section \ref{subsec:fragmentation} for a discussion on this.

The dust masses quoted in Table \ref{table: masses} are for simulations where we used the planetesimal formation criteria from \citet{Yang2017}. In section \ref{Appendix: compare SI} where we compare this criteria to the mid-plane model, it was shown that the mid-plane model produces less planetesimals and results in a larger amount of dust and pebbles remaining in the rings. More precisely, after 300,000 years the amount of dust and pebbles remaining in the ring at the innermost planet is roughly $9\, \textrm{M}_{\oplus}\, \textrm{au}^{-1}$, and the corresponding amount at the middle planet is $3\, \textrm{M}_{\oplus}\, \textrm{au}^{-1}$. For the nominal model after 300,000 years the amount of dust and pebbles remaining in the same rings is roughly $0.6-0.7\, \textrm{M}_{\oplus}\, \textrm{au}^{-1}$. Although the values from the mid-plane model appear to be a better match to the estimates by \citet{Dullemond2018}, the mid-plane criterion for planetesimal formation has not been confirmed by any hydrodynamical simulation that we know of. Since a more detailed comparison of the two planetesimal formation models is beyond the scope of this paper, the rest of the paper will only concern simulations done with our nominal model for planetesimal formation.

\subsection{Global dust distribution} \label{subsec: global dust dist}
Depending on what planet and disc parameters that are used we end up with very different pebble distributions across the disc. For simulations with a maximum grain size of around one millimeter, high drift velocities and little or no dust transportation through the planetary gaps result in the interplanetary regions becoming depleted of pebbles within a few hundred thousand years. This occurs in all simulations except for the ones with planetary masses lower than the pebble isolation mass. Combined with efficient planetesimal formation in the pressure bumps, the region interior to the outermost planet becomes devoid of pebbles (see Figure \ref{fig:discImage_all}). Such discs with large central cavities resemble transition discs \citep[e.g.][]{Andrews2011}. However, we want to emphasize that we only show the dust-to-gas surface density ratios in this work. We have not looked into how these discs would actually appear in observations of millimeter continuum emission. 

If the planetesimal formation efficiency in nature is lower than assumed in our simulations, so that a significant fraction of millimeter-sized pebbles remain in the pressure bumps, then the discs instead evolve a few very narrow and bright rings, similar to the structure observed in the protoplanetary disc around AS 209 \citep{Fedele2018}. This can be seen in Figure \ref{fig:discImage_noPlan} and Figure \ref{fig:discImage_noPlan_100micro} where we have presented simulations with no planetesimal formation and no dependency on the pressure gradient. Such discs with a high pebble density in the rings could also be created and maintained through cycles of planetesimal formation and planetesimal destruction and/or efficient fragmentation in the rings \citep[see][]{Drazkowska2019}.

Most observed protoplanetary discs with dust rings nevertheless do not appear as AS 209; instead they have emission coming more evenly from the whole protoplanetary disc (e.g. \citealt{ALMA2015}). Such dust distributions could only be obtained in our simulations by using planetary masses lower than the pebble isolation mass. Generally, if particles are transported through the planetary gaps efficiently, then the regions between the planets are continuously replenished as long as there is a large enough repository of solids far out in the disc. In our nominal simulation the outer disc still holds around 100 Earth masses of solids after $1\, \textrm{Myr}$. If dust transport through the gaps is efficient it further reduces the solid-to-gas ratios in the pressure bumps, leading to less planetesimal formation. Possible reasons for why dust transport through planetary gaps could be more efficient than in our simulations are discussed in Section \ref{subsec:destruction planetesimals}-\ref{subsec:dust filtration 1D vs 2D}. 

Introducing a maximum grain size of $100\, \mu \textrm{m}$ results in the interplanetary regions becoming depleted of their pebble mass on a longer time-scale of up to 1 million years. In these simulations one ring of pebbles remains visible inside the cavity after $1\, \textrm{Myr}$ even for planetary masses larger than the pebble isolation mass. As a comparison, the stars in the DSHARP survey have ages between a few hundred thousand to 10 million years \citep{Andrews2018}, so we know that at least some discs must be able to maintain a high pebble density in the rings for a long time. 

\subsection{Solar System constraints on planetesimal formation} \label{subsec: solar system constraints}
Since the planetesimals in our model form at the edges of planetary gaps, there must have existed an earlier population of planetesimals which participated in the formation of these gap-opening planets. Those planetesimals should have formed by some other mechanism than the one proposed in our work, e.g. through particle pile-ups outside snow lines \citep{DrazkowskaAlibert2017,SchoonenbergOrmel2017}. Our planetesimals would thus represent a second generation of planetesimals which form only once the first gap-opening planet has appeared in the disc, a scenario that fits in well with solar system observations.

In \citet{Kruijer2017} they used isotope measurements of iron meteorites together with thermal modelling of bodies internally heated by $^{26}$Al decay to study the formation of planetesimals in the asteroid belt. From this study they concluded that the parent bodies of non-carbonaceous (NC) and carbonaceous (CC) iron meteorites: 1) accreted at different times, within $0.4$ respective $0.9\,$Myr after solar system formation; 2) accreted at different locations in the disc, with the CC meteorites accreting further out; 3) must have remained separated from before $0.9$ until $3-4\,$Myr after solar system formation. This picture could be neatly explained by formation of the CC iron meteorite parent bodies at the edge of Jupiter's gap.

The first population of planetesimals form early before $0.4\,$Myr (the NC iron meteorites). Then at some time before $0.9\,$Myr Jupiter reaches the pebble isolation mass and shuts off the flow of pebbles to the inner solar system. The pressure maximum generated at the edge of Jupiter's gap promotes planetesimal formation and results in a second generation of planetesimals (the CC iron meteorites). Once Jupiter has reached the pebble isolation mass it continues to grow slowly for a few million years, keeping the NC and CC population separate. Then at $3-4\,$Myr after solar system formation something occurs that scatters the population of CC iron meteorites towards the inner Solar System and causes them to mix with the NC population. This could be the onset of runaway gas accretion \citep{Kruijer2017}, or interactions with an outer giant planet \citep{Ronnet2018}. 

\section{Shortcomings of the model}\label{setc: shortcomings of the model}

\subsection{A proper handle on fragmentation}\label{subsec:fragmentation}
We have used a collision algorithm where target particles are reduced to the mass of the projectiles in the event of a destructive collision. This is not fully realistic because destructive collisions should result in the formation of multiple fragments, and it prevents us from recovering very small particle sizes. \citet{BlumMunch1993} showed that when two similar-sized dust particles collide at a velocity higher than the fragmentation threshold, both particles are disrupted into a power-law size distribution. A significant fraction of the mass in such a collision becomes concentrated in the largest particle sizes (\citealt{Birnstiel2011,BukhariSyed2017}). Since most mass is supposed to be tied up in the largest particle sizes, our simplified algorithm is still justified, but it prevents the formation of small dust particles that could make it past the planetary gap. These particles could recoagulate interior to the gap and result in a population of millimeter-sized particles in the interplanetary region.

\citet{Drazkowska2019} used an advanced 2-D coagulation model to study the dust evolution in a disc that is being perturbed by a Jupiter-mass planet. They found that fragmentation at the gap edge indeed is important. In their simulations large grains that are trapped inside the pressure bump fragment and replenish the population of dust, which can then pass through the planetary gaps. This process will thus lead to a continuous dust flux through the gaps. Since this process removes solids from the pressure bump it should also result in less planetesimal formation; however, a comparison between the timescales for drift, fragmentation and planetesimal formation would be required in order to further assess this.

\subsection{Destruction of planetesimals}\label{subsec:destruction planetesimals}
In our simulations we only look at where planetesimals form and how much mass is turned into planetesimals. This means that we do not investigate what happens to the planetesimals once they have formed. Processes which are likely to be important in determining the fate of planetesimals at the edge of planetary gaps are: dynamical interactions with the planet, dynamical interactions with other planetesimals, and sublimation and erosion due to the flow of gas. These processes will be the subject of a follow-up study.

If the planetesimals are not removed from the gap edge directly after formation, then the density of planetesimals in this region should become very high. In such regions planetesimal collisions are likely to be frequent, and like in debris discs such events result in the production of dust \citep{Wyatt2008}. Another process that could result in a replenishing of the dust population in the pressure bumps is planetesimal sublimation due to bow shocks \citep{Tanaka2013}. The heating and sublimation of the planetesimals result in a shrinking of the planetesimal size, and the vapour can form dust particles through recondensation. How efficient and relevant these processes are for the production of dust in a pressure bump remains to be studied. 

\subsection{The streaming instability may not be 100\% efficient}\label{subsec: efficiency of SI}
We assume in our model that whenever the critical density to trigger the streaming instability is reached, planetesimals form. The planetesimal formation algorithm used in this study results in a maximum efficiency for planetesimal formation. A calculation of the actual formation efficiency would require taking into account the timescale for collapse into planetesimals, the timescale for particles to drift across the gap, as well as turbulence and many other effects. However, in simulations where there is no transport of pebbles across the gaps the efficiency should not play a big role. It does not matter if it takes a hundred years or a hundred thousand years to form planetesimals since the particles will anyway remain trapped. In simulations where pebbles are able to make it past the gaps, like in the simulations with 100 micron sized particles, the efficiency for planetesimal formation becomes much more important. 

One mechanism which would likely result in less planetesimal formation is efficient fragmentation in the pressure bump, which was discussed in section \ref{subsec:fragmentation}. We also stress that the linear scaling with the pressure gradient is an approximation, and if this relation was less steep or levelled out towards a pressure gradient of zero, more pebbles and dust would be left in the pressure bumps. Further, in 1-D simulations we do not need to worry about instabilities at the gap edges. However, if the gaps are deep enough in 2-D or 3-D simulations the gap edges may become unstable to form vortices \citep{HallamPaardekooper2017}. The triggering of vortices could potentially change the efficiency of planetesimal formation; however, planetesimal formation in such an environment is still poorly understood. Finally, the coagulation model from \citet{guttler2010} results in a bimodal particle size distribution. In \citet{Krapp2019} it was shown that the streaming instability becomes less efficient when there are multiple particle sizes involved. However, the difference in growth rate between single particles and multiple species appears to vanish when the dust-to-gas ratio is above unity. Therefore, we continue to use the mass averaged Stokes number in our planetesimal formation model, and do not lower the efficiency when the particle size distribution evolves into bimodal. 

\subsection{Dust filtration through planetary gaps in 1D versus 2D simulations}\label{subsec:dust filtration 1D vs 2D}
There could be a systematic difference in the dust filtration by a planet in 2-D simulations relative to our 1-D simulations. This was shown by \citet{Weber2018} and \citet{Haugbolle2019} who performed detailed studies of dust filtration through planetary gaps. In these works they used a dust fluid approach in order to track extremely low values of the dust density. They found that dust is more likely to be transported through the gaps in 2-D simulations, although the amount on the interior of the planet orbit is diminished greatly by the filtering. In \citet{Drazkowska2019} they instead found the opposite results when comparing dust filtration in 1-D and 2-D coagulation simulations. One possible reason for this discrepancy could be that the gap profiles in \citet{Drazkowska2019} were the same in both the 1-D and 2-D simulations, while in \citet{Weber2018} the density profiles varied between the 1-D and 2-D simulations. Regardless, it seems clear that dust filtration in 1-D simulations do differ from the 2-D or 3-D case; however, exactly how is still not certain.

\subsection{The $\alpha$-disc model}
In the classical $\alpha$-disc model a macroscopic viscosity is assumed to drive angular momentum transport throughout the disc. However, the actual origin of this viscosity is not known. Alternatively, the angular momentum may be drained from the protoplanetary disc by strong winds. In such models mass is primarily removed from the disc surface, and not from the mid-plane. The surface density profile of such wind-driven discs vary a lot in the literature, and while some resemble the classical $\alpha$-disc model, others have positive density gradients in the inner regions of the disc and multiple density maxima spread across the disc (\citealt{Gressel2015,Bai2016,Suzuki2016,Bethune2017,Hu2019}). In such discs particle drift would be very different from what we use in our model, and our results would therefore change. However we still do not know enough about what drives angular momentum transport in discs, or what the level of turbulent viscosity and wind transport are, to say anything for sure about which model is correct. Therefore we decided here to stick to the well-understood $\alpha$-model.

\section{Conclusions and future studies}\label{setc: conclusion}
In this work we test the hypothesis that dust trapping at the edges of planetary gaps can lead to planetesimal formation via the streaming instability. To study this we perform 1-D global simulations of dust evolution and planetesimal formation in a protoplanetary disc that is being perturbed by multiple planets. We perform a parameter study to investigate how different particle sizes, disc parameters and planetary masses affect the efficiency of planetesimal formation. We further compare the simulated pebbles’ distribution with protoplanetary disc observations. 

The answers we have obtained for the questions posed in the introduction can be summarized as follows:

\begin{itemize}
\item[1.] \textit{Do planetesimals form at the edges of planetary gaps}?\\
Planetesimal formation occurs in all of our simulations and is almost exclusively limited to the edges of planetary gaps.  
\item[2.] \textit{How efficient is this process and how does the efficiency vary with different disc and planet parameters}?\\
Planets with masses above the pebble isolation mass trap pebbles efficiently, and in the case of millimeter-sized particles essentially all of these trapped pebbles are converted into planetesimals. As long as the pebbles can not pass through the gaps, the amount of planetesimals that form does not vary between the simulations, although the region in which they form do change a bit. Decreasing the pebble size to 100 micron results in less efficient conversion of pebbles to planetesimals and more transport through the gaps. 
\item[3.] \textit{What does the distribution of dust and pebbles look like for the different simulations}?\\
In the case of millimeter-sized pebbles and planetary masses larger than the pebble isolation mass the region interior to the outermost planet gets depleted of pebbles in a few hundred thousand years. For planetary masses lower than this, transport through the gaps leads to a constant replenishment of the interplanetary region, resulting in a gap-and-ring like pebble distribution. In the case where the particle size was lowered to 100 $\mu$m, there is always at least one ring of pebbles remaining inside the cavity. When we lower the efficiency of planetesimal formation, by ignoring the drop in the metallicity threshold for planetesimal formation with decreasing pressure support, the discs instead appear to have narrow and bright rings.
\item[4.] \textit{How do these distributions compare with observations of protoplanetary discs}?\\
Transition discs with large central cavities are known from observations. Similar discs with large cavities are the general outcome of simulations with massive planets, millimeter-sized pebbles, and efficient planetesimal formation. Discs with narrow and bright rings, similar to the outer regions of the disc around the young star AS 209, are the outcome of simulations with massive planets but low planetesimal formation efficiency in the rings. A replenishment of the dust population in the rings, through processes such as fragmentation, planetesimal collisions or planetesimal evaporation and erosion, could result in similar structures and potentially also aid in transporting particles across the gaps \citep{Drazkowska2019}. Setting the maximum grain size to 100 $\mu$m results in multiple rings, longer drift time-scales and a larger variety in disc structures. Generally, the only simulations that could produce images similar to HL Tau, with multiple gap and ring structures but no strong pebble depletion anywhere in the disc, are simulations with planetary masses lower than the pebble isolation mass.  
\end{itemize}

In this work we have focused on studying the efficiency of planetesimal formation and the locations of the formed planetesimal belts, but we have neglected their further evolution after formation. Processes which are likely to be important in determining the fate of planetesimals formed at the edges of planetary gaps are: dynamical interactions with the gap-forming planets, planetesimal-planetesimal interactions, and erosion and evaporation by the flow of gas. These processes will be the subject of a follow-up study.

\begin{acknowledgements}
The authors wish to thank Alexander J. Mustill and the anonymous referee for helpful comments that led to an improved manuscript. The authors further wish to thank Alessandro Morbidelli for inspiring discussions, and Chao-Chin Yang for comments and advice regarding the planetesimal formation model. L.E., A.J. and B.L. are supported by the European Research Council (ERC Consolidator Grant 724687-PLANETESYS). A.J. further thanks the Knut and Alice Wallenberg Foundation (Wallenberg Academy Fellow Grant 2017.0287) and the Swedish Research Council (Project Grant 2018-04867) for research support. The computations were performed on resources provided by the Swedish Infrastructure for Computing (SNIC) at the LUNARC-Centre in Lund, and are partially funded by the Royal Physiographic Society of Lund through grants.
\end{acknowledgements}

\bibliographystyle{aa} 
\bibliography{ref} 

\appendix

\section{Particle collisions}\label{Appendix: collision algorithm}
Particle collisions are done through a Monte Carlo method. Each particle swarm is assigned a total mass $M_i$, an individual particle mass $m_i$, and a number density $n_i=M_i/m_i$. The total mass $M_i$ is different for different particles, which is useful in order to resolve a wide range of column densities in the disc. 

When two particles collide we define the larger particle as the target and the smaller particle as the projectile. The rate of interaction (defined below) between the target and the projectile is determined as
\begin{equation}\label{eq: mass doubling rate 1}
r_{tp}=  \sigma_{tp} v_{tp} n_p \frac{m_p}{m_t} \max(M_t/M_p,1),
\end{equation}
where $\sigma_{tp}$ and $v_{tp}$ are the collisional cross section and relative speed between the target ($t$) and the projectile ($p$). If the total mass in the projectile is larger than or equal to the total mass in the target, then the interaction time-scale is defined as the time for each target particle in a swarm to collide with its own mass in projectiles. If the total mass in target particles is larger than the total mass in projectiles, we multiply by $M_t/M_p$ so that the interaction time-scale is instead the time-scale for all projectile particles to have collided with a target particle.

Equation \ref{eq: mass doubling rate 1} can be rewritten as 
\begin{equation}
\begin{split}
r_{tp} &=  \sigma_{tp} v_{tp} n_t \frac{M_p}{M_t} \max(M_t/M_p,1) \\
&=  \sigma_{tp} v_{tp} n_t \max(M_p/M_t,1).
\end{split}
\end{equation}
Using that the collisional cross section is $\sigma_{tp}=\pi (s_t+s_p)^2$, we obtain the final equation for the mass doubling rate
\begin{equation}
r_{tp}=\pi (s_t+s_p)^2 v_{jk} n_t \max(M_p/M_t,1),
\end{equation}
where $s$ is particle radius. The relative speed contains contributions from Brownian motion, differential radial and azimuthal drift, differential reaction to the gas accretion speed, and turbulent speed. The turbulent speed is based on the closed-form expressions of \citet{OrmelCuzzi2007}, all other terms are standard in the literature \citep[see e.g.][]{Brauer2008}.

For averaging over the vertical direction, we assume that the two particle species maintain a Gaussian density profile in the vertical direction and that changes to the target particle by coagulation are immediately diffused over the entire column density of the target particle. We follow here a similar approach to \citep{Brauer2008}, their Appendix B. The coagulation equation written for a given height $z$ over the mid-plane is
\begin{equation}\label{eq: coagulation eq}
r_{tp}(z) = \pi (s_t+s_p)^2 v_{tp} n_p(z).
\end{equation}
The collision rate averaged over $n_t(z)$ is then
\begin{equation}
\overline{r_{tp}} = \frac{\int r_{tp}(z) n_t(z) dz}{\int n_t(z) dz}.
\end{equation}
Assuming a Gaussian density distribution with mid-plane number density $n_0,{\rm t}$ and scale-heights $H_{\rm t}$ and $H_{\rm p}$, the integration yields
\begin{equation}
\overline{r_{tp}} = \pi (s_t+s_p)^2 v_{tp} n_{0,i} \times \frac{1}{\sqrt{1+(H_t/H_p)^2}}.
\end{equation}
Comparing to equation \ref{eq: coagulation eq} we see that the vertical integration of the coagulation equation can be treated as a simple multiplication factor on the rate of collisions in the mid-plane.

In order to calculate the number density of particles in the mid-plane, we divide the mass in each superparticle by the area of the annulus where the particle is present ($2 \pi r \Delta r$) and then by $1/[\sqrt{2\pi}H_p]$ to obtain the mid-plane density.

The time-step contribution from particle coagulation is based on the interaction rate $r_{ij}$. The time-step for a particle $i$ is  
\begin{equation}
\tau_i=\frac{1}{\sum_j(r_{ij})}. 
\end{equation}
The Monte Carlo time-step is then $\min_i(\tau_i)$ times a numerical factor, chosen to be $0.2$. Once the time-step has been calculated, we loop over all the particles in a grid cell and all their unique partners. For each particle pair we draw a random number, and if that number is smaller than $dt\times r_{ij}$, we let the swarms interact. We base the outcome of collisions on experimental results by G{\"u}ttler et al. (2010), and we assume that the particles are porous. The possible outcomes of a collision are sticking, bouncing, bouncing with mass transfer, and fragmentation. Sticking means that the target either doubles its mass or multiplies its mass by $(1+M_p/M_t)$, if $M_p<M_t$. For bouncing with mass transfer we double the mass of the projectile particles, and subtract the projectile particle mass from each target particle. For fragmentation we set all the target and projectile particles to the mass of the projectile. If there are excess target particles, then they retain their original mass.

\section{Particle size distributions}\label{Appendix: sec size dist}
\begin{figure*}
\resizebox{\hsize}{!}
    {\includegraphics{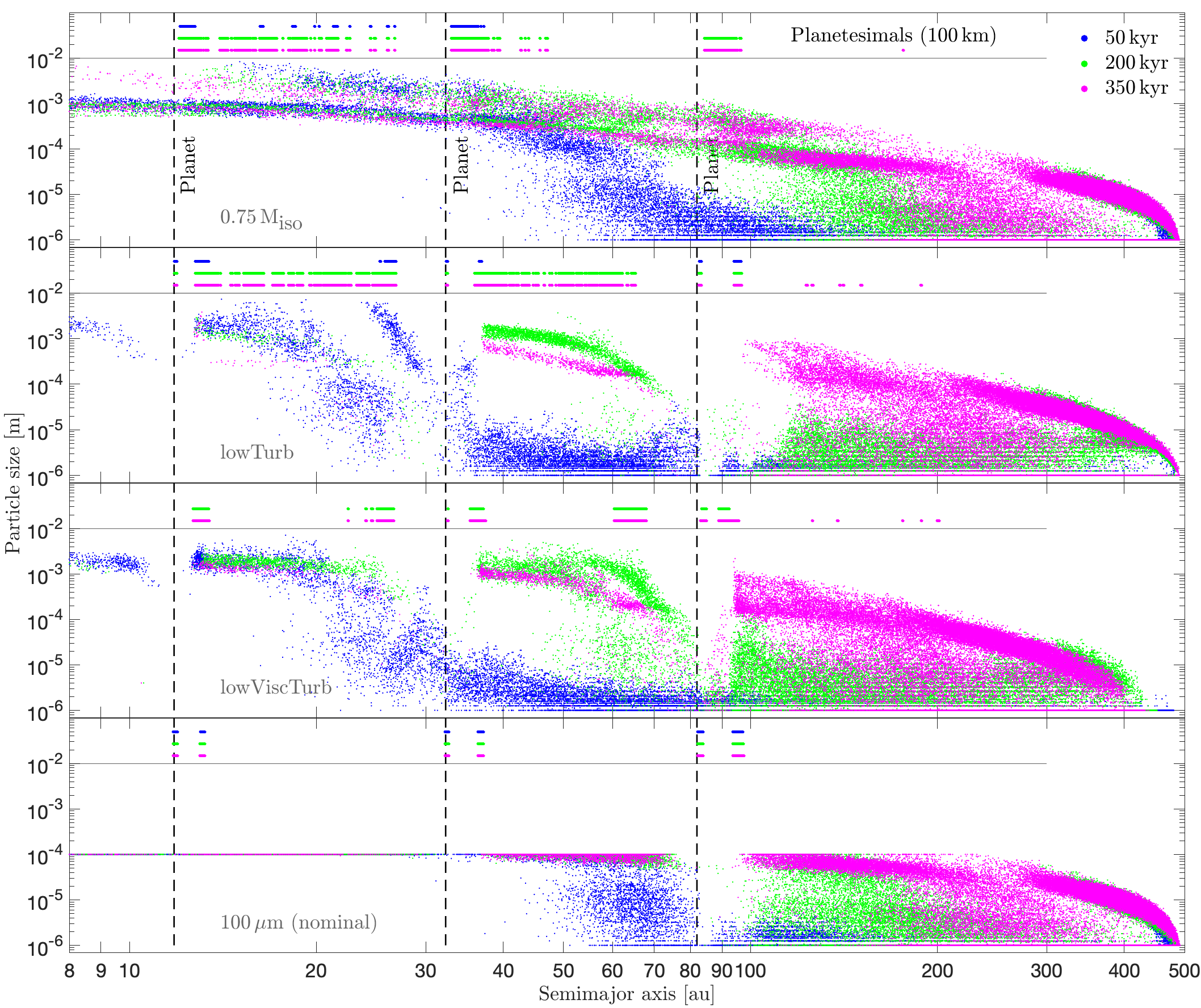}}
\caption{Particle size distributions at different times during disc evolution for four different simulations. The semimajor axes of the formed planetesimals are indicated at the top of the plots. Top panel: in the simulation with a planetary mass of $0.75\, \textrm{M}_{\textrm{iso}}$, planetesimals form in a wider region around the pressure bump than in the nominal simulation. Less efficient pebble trapping at the pressure bump also results in a more even distribution of dust and pebbles in the disc, with no strong depletion at the location of the planets. Second panel: lowering the turbulence diffusion to $10^{-4}$ results in slower collisional velocities, which results in slower coagulation, but eventually leads to larger particle sizes. Third panel: when the viscous parameter is lowered to $10^{-4}$ as well, we get small bumps in the gas surface density profile at the inner edges of the planetary gaps. Particles become trapped in these bumps, resulting in some planetesimal formation also at these locations. Bottom panel: this plot shows the implementation of a maximum grain size of $100\, \mu$m. }
    \label{Appendix: fig size dist}
\end{figure*}

Figure \ref{Appendix: fig size dist} shows the size distributions of particles at different times during disc evolution for some selected simulations. When a planetary mass is used that is lower than one pebble isolation mass (top panel), the gaps never get completely depleted of dust and pebbles. The pile-up of material at the gap edges is also much less prominent, and since pebbles are now transported through the planetary gaps, the result is a more even distribution of particles throughout the disc. Since there are more particles in the interplanetary regions, we also get more spontaneous concentrations leading to a planetesimal formation. A comparison with Figure \ref{fig:hist_all} shows that the amount of planetesimals forming in the interplanetary regions is still negligible compared to the amount that forms at the gap edges. 

When the turbulent diffusion is lowered by an order of magnitude to $10^{-4}$ (second panel), the coagulation time-scale increases. Since it takes more time for particles to grow, it also takes more time for the size distribution to become bimodal. The slow particle growth also results in that the drift time-scale is initially longer. However, decreasing the amount of turbulence results in lower collisional speeds, which in turn results in larger particle sizes. For such large particles, the time-scale for drift becomes shorter than in the nominal model. The result is that more particles make it from the exponentially tapered outer disc to the inner 100 au where the planets reside. The larger particle sizes also leads to that sporadic concentrations become more common, again resulting in more planetesimal formation in the interplanetary regions. 

In the simulation where both the viscous parameter and the turbulence diffusion has been decreased to $10^{-4}$ (second panel), trapping at the inner gap edges results in a significant amount of planetesimals being formed at these locations. A small bump in the gas surface density profile is created at the beginning of all simulations, but when the viscous parameter is high this bump disappears before planetesimal formation is initiated. For a viscous parameter of $10^{-4}$, this pile-up of gas at the inner gap edges is both more prominent and longer lasting than in all other simulations. Since the viscosity is small, the time for gap-clearing is also longer. 

In the bottom panel of Figure \ref{Appendix: fig size dist} we show the size distribution for particles in the nominal model when a maximum grain size of $100\, \mu$m has been applied. Far out in the disc this constraint does not matter for the particle evolution, since particles does not grow that large anyway. 

\end{document}